\RequirePackage{fix-cm}
\RequirePackage{fixltx2e}
\pdfoutput=1
\documentclass[aps,pre,twocolumn,superscriptaddress,10pt,nofootinbib,showpacs]{revtex4-1}

\usepackage{subfigure}
\usepackage{amsfonts, amssymb, amsmath}
\usepackage{bm}
\usepackage{graphics}
\usepackage{graphicx}
\usepackage{wasysym}
\usepackage{natbib}
\usepackage{mathtools}
\usepackage{placeins}

\usepackage[utf8x]{inputenc}
\usepackage[T1]{fontenc}

\usepackage[usenames,dvipsnames]{xcolor}

\usepackage{color}
\definecolor{darkblue}{rgb}{0,0,0.6}
\definecolor{darkred}{rgb}{0.6,0,0}

\usepackage{tikz}
\usepackage{hyperref}
\hypersetup{
bookmarksopen=true,
pdftitle= KPZ equation with short-range correlated noise: emergent symmetries and non-universal observables,
pdfauthor= Steven Mathey et al.,
pdftoolbar=false,
pdfstartview={FitH},
pdfmenubar=true,
pdfhighlight=/O,
colorlinks=true,
urlcolor=darkblue,
citecolor=darkblue,
linkcolor=MidnightBlue,
}

\newcommand{\moy}[1]{\left\langle  #1 \right\rangle }

\renewcommand{\vec}[1]{{\boldsymbol{#1}}}
\newcommand{\tx}{{t,\vec{x}}}
\newcommand{\txp}{{t',\vec{x}'}}
\newcommand{\op}{{\omega,\vec{p}}}
\newcommand{\mop}{{-\omega,-\vec{p}}}
\newcommand{\I}{\text{i}}
\newcommand{\taur}{{\tau,\vec{r}}}
\newcommand{\ie}{i.e.~}

\newcommand{\eg}{e.g.~}

\newcommand{\sref}[1]{Sec.~\ref{#1}}
\newcommand{\fref}[1]{Fig.~\ref{#1}}
\newcommand{\tref}[1]{Tab.~\ref{#1}}
\newcommand{\aref}[1]{Appendix~\ref{#1}}

\newcommand{\Eq}[1]{Eq.~(\ref{#1})}
\newcommand{\eq}[1]{(\ref{#1})}
\newcommand{\Eqs}[1]{Eqs.~(\ref{#1})}

\usepackage{abbrevs}
\usepackage{etoolbox}
\newabbrev\RG{Renormalization Group (RG)}[RG]
\newabbrev\FRG{Non-perturbative Functional Renormalization Group (NP-FRG)}[NP-FRG]
\newabbrev\KPZ{Kardar--Parisi--Zhang (KPZ)}[KPZ]
\newabbrev\IR{Infrared (IR)}[IR]
\newabbrev\UV{Ultraviolet (UV)}[UV]
\newabbrev\BMW{Blaizot--Mendez--Wschebor (BMW)}[BMW]
\newabbrev\MSR{Martin--Siggia--Rose Janssen--de Dominicis (MSRJD)}[MSRJD]
\newabbrev\NLO{Next-to-Leading Order (NLO)}[NLO]
\newabbrev\SO{Second Order (SO)}[SO]
\newabbrev\GVM{Gaussian Variational Method (GVM)}[GVM]
\newabbrev\OneD{one-dimensional ($1d$)}[$1d$]
\newabbrev\DP{Directed Polymer (DP)}[DP]

\robustify{\RG}
\robustify{\FRG}
\robustify{\KPZ}
\robustify{\UV}
\robustify{\IR}
\robustify{\BMW}
\robustify{\MSR}
\robustify{\NLO}
\robustify{\SO}
\robustify{\GVM}
\robustify{\OneD}
\robustify{\DP}

\makeatletter
\renewcommand\maybe@space@{%
  \maybe@ictrue % <= this is new
  \expandafter   \@tfor
    \expandafter \reserved@a
    \expandafter :%
    \expandafter =%
                 \nospacelist
                 \do \t@st@ic
  \ifmaybe@ic % <= this is new
    \space
  \fi
}
\makeatother

\begin{document}

\title{KPZ equation with short-range correlated noise: emergent symmetries and non-universal observables}

\author{Steven Mathey}
\email{steven.mathey@lpmmc.cnrs.fr}
\affiliation{LPMMC, Universit\'e Grenoble Alpes and CNRS, F-38042 Grenoble, France}

\author{Elisabeth Agoritsas}
\affiliation{LIPhy, Universit\'e Grenoble Alpes and CNRS, F-38042 Grenoble, France}
\affiliation{Laboratoire de Physique Th\'{e}orique, ENS \& PSL University, UPMC \& Sorbonne Universit\'{e}s, CNRS, 75005 Paris, France}

\author{Thomas Kloss}
\affiliation{INAC-PHELIQS, Université Grenoble Alpes and CEA, 38000 Grenoble, France}

\author{Vivien Lecomte}
\affiliation{LIPhy, Universit\'e Grenoble Alpes and CNRS, F-38042 Grenoble, France}
\affiliation{LPMA, Universit\'e Paris Diderot and Pierre et Marie Curie and CNRS, F-75013 Paris, France}

\author{L\'{e}onie Canet}
\affiliation{LPMMC, Universit\'e Grenoble Alpes and CNRS, F-38042 Grenoble, France}

\date{\today}

\begin{abstract}
We investigate the stationary-state fluctuations of a growing one-dimensional interface described by the \KPZ dynamics with a noise featuring smooth spatial correlations of characteristic range $\xi$. We employ Non-perturbative Functional Renormalization Group methods in order to resolve the properties of the system at all scales. We show that the physics of the standard (uncorrelated) \KPZ equation emerges on large scales independently of $\xi$. Moreover, the Renormalization Group flow is followed from the initial condition to the fixed point, that is from the microscopic dynamics to the large-distance properties. This provides access to the small-scale features (and their dependence on the details of the noise correlations) as well as to the universal large-scale physics. In particular, we compute the kinetic energy spectrum of the stationary state as well as its non-universal amplitude. The latter is experimentally accessible by measurements at large scales and retains a signature of the microscopic noise correlations. Our results are compared to previous analytical and numerical results from independent approaches. They are in agreement with direct numerical simulations for the kinetic energy spectrum as well as with the prediction, obtained with the replica trick by Gaussian variational method, of a crossover in $\xi$ of the non-universal amplitude of this spectrum.
\end{abstract}


\maketitle

\section{Introduction}
\label{section-intro}
\ResetAbbrevs{All}
 
Introduced three decades ago \cite{Kardar1986a}, the \KPZ equation is one of the simplest non-linear Langevin equations. As such, it relates generically to a broad range of systems, within the so-called \KPZ universality class; see \cite{HalpinHealy1995,Kriecherbauer2010,Sasamoto2010a,Corwin2012a,quastel_lecture-notes-arizona_2012,Takeuchi2014,Quastel2015,Halpin-Healy2015,Sasamoto2016} and references therein. Thus, stochastic growth of roughening interfaces, for which the \KPZ equation was initially introduced \cite{Kardar1986a,krug_1997_AdvPhys_46_139}, shares common features with systems as dissimilar as, for instance, the Burgers equation in hydrodynamics \cite{book-burgers,Bec2007a}, \DP in random media \cite{huse_henley_fisher_1985_PhysRevLett55_2924,bouchaud_mezard_parisi_1995_PhysRevE52_3656}, random matrices \cite{johansson_2000_CommMathPhys209_437,praehofer_spohn_2000_PhysRevLett84_4882}, the dynamics of Bose gases \cite{Kulkarni2013,Gladilin2014,mathey2014thesis,Ji2014,Altman2013a,mathey2014,He2014a,Kulkarni2015,Mendl2016} or active fluids \cite{Chen2016b}. The deep connections within the \KPZ universality class allow, on the one hand, to provide known results with alternative physical interpretations in different languages (turbulence in hydrodynamics, \DP free-energy landscape, Fredholm determinants of random matrices, etc.) and, on the other hand, to successfully export these results from one problem to another.

In particular, in its original formulation, the \KPZ equation is a stochastic continuum equation with an \emph{uncorrelated} noise and, although remarkably difficult to address exactly, its \OneD fluctuations and universal features have been recently completely elucidated for the stationary state \cite{huse_henley_fisher_1985_PhysRevLett55_2924,halpin-healy_diverse_1989}, the approach to this stationary state \cite{dotsenko_bethe_2010,calabrese_free-energy_2010,sasamoto_one-dimensional_2010,amir_probability_2011} and even the complete time-dependence, starting from different initial conditions (\ie flat, `sharp-wedge', or stochastic) \cite{Calabrese2011,le_doussal_kpz_2012,gueudre_directed_2012,imamura_exact_2012}.
However, when  the original \KPZ model is slightly modified (see \cite{Amar1991b,Aranson1998,kloss_2014_PhysRevE90_062133,Strack2014a,kloss_2014_PhysRevE89_022108,Gueudre2015,Sieberer2016} and references therein for examples), these exact solutions are in general no longer valid, and a key open issue is to assess what the robustness of the \KPZ universal features is, and to determine under what conditions they can be expected to persist.

The present study focuses specifically on the role of a \emph{spatially-correlated} noise, on a characteristic length ${\xi>0}$. Note that a noise with spatial power-law correlations was studied numerically \cite{Peng1991,Hayot1996a,Li1997,Verma2000,Chu2016} and analytically \cite{Medina1989a,Janssen1999a,frey1999scaling,kloss_2014_PhysRevE89_022108}. In contrast, we explicitly include a correlation length here. Such an ingredient is crucial physically, since in experimental systems $\xi$ is always finite. Nevertheless, if universality truly holds, then the macroscopic scale invariance is expected to be independent of this microscopic correlation length and thus to be captured by the same \RG fixed point as for the uncorrelated case (${\xi=0}$). In particular, we focus on a \OneD interface in the asymptotic stationary state at long times where we fully resolve the dependence of the two-point correlation function on the relative space and time [see \Eq{eq:def_c}]. We show that, although the universal features of the original \KPZ equation with $\delta$-correlated noise ($\xi=0$) emerge on large scales, the small-scale physics is strongly dependent on $\xi$ and is thus, non-universal.

Experimentally, the \KPZ dynamics was realised in a wide range of platforms. To give a few examples, the universal \KPZ physics was observed in fronts of slow combustion of paper \cite{Miettinen2005}, at the separation between flat crystal facets and rounded edges \cite{Degwa2006}, at the interface of different modes of turbulence in liquid crystals \cite{Takeuchi2012a}, in patterns in the deposition of particles suspended in evaporating droplets \cite{yunker2013}, in chemical reaction fronts in disordered media \cite{Atis2014}, in the shape of growing colonies of bacteria and cells \cite{Muzzio2014,huergo2015} and, most recently, in epitaxy-driven film growth \cite{Almeida2014,halpin-healy2014,Almeida2015}. See \eg \cite{Takeuchi2014} or Sec.~VII. of \cite{agoritsas_2012_FHHtri-analytics} for overviews and further references. In this article we discuss in particular the non-universal amplitude of the interface roughness [see \Eq{eq:roughness_dtilde}] which is a large-scale observable that could be accessible experimentally even though it depends explicitly on the microscopic correlation length. See sections \ref{section-discussion-Dtilde} and \ref{section-connection-exp} for details.

The role of a finite $\xi$ has been addressed in a series of previous studies in the language of the $1+1$ \DP endpoint \cite{agoritsas_2010_PhysRevB_82_184207,agoritsas_2012_ECRYS2011,agoritsas-2012-FHHpenta,agoritsas_2012_FHHtri-analytics,agoritsas_2012_FHHtri-numerics,phdthesis_Agoritsas2013} and recently in \cite{agoritsas_lecomte_2016_scalings-GVM,Dotsenko2016}, combining analytical and numerical approaches to characterise the complete time dependence of the \KPZ fluctuations, starting from the so-called `sharp-wedge' initial condition. It was predicted in particular (within the \GVM and a Bethe ansatz analysis as well as with a direct numerical integration of the correlated \KPZ equation) that the amplitude of the \KPZ fluctuations must display a crossover, as $\xi$ is tuned, from the exact solution of the uncorrelated case ($\xi=0$) \cite{Sasamoto2010a,Calabrese2011,Corwin2012a} to a regime of large-scale correlations. Still no exact analytical expression has been obtained so far, not even for the stationary state. When no exact solution is available ($d>1$, correlated noise, non-Gaussian noise, etc.) non-perturbative approaches are necessary to study the \KPZ physics. Here we employ a very versatile method, the \FRG \cite{Wetterich:1992yh}  (for reviews, see \cite{Bagnuls:2000ae,Berges:2000ew,Polonyi:2001se,Kopietz10,Delamotte:2007pf}), to investigate the universal versus non-universal features of the stationary two-point correlation function, in the presence of a finite correlation length\footnote{Note that the \RG approach that we use is complementary to perturbative functional renormalization schemes. See \eg \cite{ledoussal_2008_arXiv:0809.1192}.}.

The \FRG method has been used in a very broad range of problems, from high- to low-energy physics. In statistical physics it has led to very accurate \cite{Canet:2003qd,Benitez2009,Benitez2012} and fully non-perturbative \cite{Canet2004a,Tarjus04,Canet2005a,Tissier06,Essafi2011,Gredat2014} results. In particular, it has been extended to the study of classical non-equilibrium systems in \cite{Canet:2003yu,canet2011b,Berges2012a,Mesterhazy2013a}. Recent successful applications include the dynamical Random Field Ising model \cite{Balog2013a,Balog2015a} or, in the case of far-from-equilibrium dynamics, fully developed turbulence \cite{Mejiamonasterio2012a,mathey2014thesis,mathey2014,Pagani2015a,Canet2014b,Canet2016} and driven-dissipative Bose gases \cite{Sieberer2015b}. Examples of its application to non-stationary dynamics can be found \eg in \cite{Gezzi2007a,Jakobs2007a,Karrasch2010,Gasenzer:2010rq,Chiocchetta2016}. The \FRG formalism to study the \KPZ equation has been developed in \cite{Canet2005b,canet_2010_PhysRevLett104_150601,canet_2011_PhysRevE84_061128,kloss_canet_wschebor_2012_PhysRevE86_051124}, where it has allowed for an accurate description of its stationary properties. In particular, the two-point correlation function obtained in this framework reproduces the exact \OneD solution with an unprecedented accuracy \cite{canet_2011_PhysRevE84_061128}. Predictions for universal ratios in $d=2$ and $3$ \cite{kloss_canet_wschebor_2012_PhysRevE86_051124} were recently tested in large-scale numerical simulations which showed a remarkable agreement \cite{HalpinHealy2013a,HalpinHealy2013b}. Several extensions of the \KPZ dynamics have also been studied in the \FRG framework, such as the presence of power-law spatially correlated noise \cite{kloss_2014_PhysRevE89_022108} (see also \cite{Strack2014a} for temporally correlated noise), and of spatial anisotropy \cite{kloss_2014_PhysRevE90_062133}. We here adapt this approach to the presence of a microscopic noise with finite range correlations and investigate both universal and non-universal features.

The outline of the paper is the following. We first set up in \sref{sec:setup_and_tools} the model centred on the \OneD \KPZ stationary-state fluctuations, the corresponding \FRG formalism, and the specific approximation scheme  considered. Then we present in \sref{section-NPRG-universal-vs-nonuniversal} our results for the two-point correlation function, discussing successively the \RG fixed point, the scaling form of the correlator, and  its non-universal features. We show that universality is recovered at sufficiently large scales, in the sense that the \RG fixed point and the scaling form turn out to be the same as for the uncorrelated noise case. On the other hand, the non-universal features depend on the specific noise correlator, and we discuss in \sref{section-connection-exp} the experimental application of our results to probe the microscopic noise correlation. We summarise and present some perspective to this work in \sref{section-conclusion-perspectives}. Additional details are gathered in the appendices.

\section{Set-up and theoretical tools}
\label{sec:setup_and_tools}

\subsection{\texorpdfstring{\KPZ}{KPZ} equation with a correlated noise}
\label{sec:kpz_definitions}

We consider the \KPZ equation \cite{Kardar1986a} in the presence of spatial correlations in the microscopic noise
\begin{align}
 \begin{split}
 & \partial_{t} h = \frac{\lambda}{2}\left[\vec{\nabla}h\right]^2 + \nu \nabla^2 h + \eta \, , \\
 & \moy{\eta(t_1,\vec{x}_1) \eta(t_2,\vec{x}_2)} = 2 D \delta(t_1-t_2) R_{\xi'}(\vec{x}_1-\vec{x}_2) \,.
 \end{split}
 \label{eq:kpz}
\end{align}
$h(\tx)$ is the time $t$, and space $\vec{x}$, dependent field that describes the interface height. $\eta(\tx)$ is a stochastic noise with Gaussian statistics and zero average, and angular brackets $\moy{\cdot}$, denote averages over $\eta$. The noise correlator $R_{\xi'}$ is an analytic function that decays to zero at a characteristic scale $\xi'$, and that is normalised as  $\int_\vec{r} R_{\xi'}(\vec{r}) = 1$. Here and in the following, we use the short-hand notation $\int_\tx = \int \text{d}^dx \text{d}t$ and $\int_\op = 1/(2\pi)^{d+1}\int \text{d}^dp \text{d}\omega$. Unless stated otherwise, $R_{\xi'}$ is taken as a Gaussian function\footnote{\label{ft:footnote_fourier}The same symbol are used for functions and their Fourier transforms, which are differentiated by their arguments ($t,x$ for real and $\omega,p$ for Fourier space).}, 
\begin{align}
 R_{\xi'}(\vec{x}) = \frac{1}{\left(2 \pi \xi'^2\right)^{\frac{d}{2}}} \, \text{e}^{- \frac{x^2}{2 \xi'^2}} \, ,
  && R_{\xi'}(\vec{p}) = \text{e}^{- \frac{\xi'^2 p^2}{2}} \, \label{eq:corr-xi},
\end{align}
where $x = \sqrt{\vec{x}^2}$ and similarly for $p$. We choose a system of units where the correlation length $\xi = \xi' D \lambda^2/\nu^3$, is the only dimensionless parameter left after rescaling, such that \Eq{eq:kpz} becomes
\begin{align}
\begin{split}
 & \partial_t h = \frac{1}{2}\left[\vec{\nabla}h\right]^2 + \nabla^2 h + \eta \, , \\
 & \moy{\eta(t_1,\vec{x}_1) \eta(t_2,\vec{x}_2)} = 2 \delta(t_1-t_2) R_\xi(\vec{x}_1-\vec{x}_2) \, .
 \end{split}
 \label{eq:kpz_du}
\end{align}

We focus on the fluctuations of the profile ${h(t,\vec{x})}$ in the stationary state, in which the average height ${\moy{h}}$ is finite and corresponds to a drift linear in time of the growing interface. The two-point correlation function in the stationary state is defined as
\begin{align}
 C(\tau,\vec{r})   & = \moy{h(\tau,\vec{r})h(0,\vec{0})} - \moy{h(\taur)}\moy{h(0,\vec{0})} \nonumber  \\
  & =\moy{h(\tau,\vec{r})h(0,\vec{0})}_c \, .
  \label{eq:def_c}
\end{align}
Using connected correlation functions, denoted by the subscript $c$, amounts to working in the co-moving frame where the average height field $\moy{h}$ is subtracted out. The correlation function $C(\taur)$ is related to the usual interface width $W(\taur)$, by
\begin{align}
 W(\taur) & = \moy{\left[h(\taur)-h(0,\vec{0})\right]^2}_c \nonumber \\
 & = 2\left[C(0,\vec{0})-C(\taur)\right] \, .
\end{align}
Time-translational invariance in space and time is assumed. Note that the time variable $\tau$ is a time difference in the stationary state. For the standard \KPZ dynamics ($\xi=0$), this correlation function exhibits scale-invariance on large spatio-temporal scales, where it takes the scaling form
\begin{align}
 C(\taur) = r^{2\chi} \, g\left(\frac{\tau}{r^z}\right) \, ,
 \label{eq:chi_and_z}
\end{align}
and $r=|\vec{r}|$. $\chi$ and $z$ are the universal roughness and dynamical critical exponents and $g$ is a universal scaling function. In \OneD, the exponents take the exact values $\chi=1/2$ and $z=3/2$.

\subsection{KPZ field theory}
\label{section-KPZ-field-theory}

The stationary state of the stochastic \KPZ equation \Eq{eq:kpz_du} is described by the generating functional (in the path integral representation)
\begin{align}
 Z[J,\tilde{J}] & = \moy{\text{e}^{\int_\tx \left( h J + \tilde{h} \tilde{J}\right)}} \nonumber \\
 & = \int D[h,\I\tilde{h}] \, \text{e}^{-S[h,\tilde{h}]+\int_\tx \left(h J + \tilde{h} \tilde{J}\right)} \, ,
 \label{eq:generating_functional}
\end{align}
where the space-time dependence of the fields inside local integrals is implicit.  $S[h,\tilde{h}]$ is the Martin--Siggia--Rose Janssen--de Dominicis action \cite{Martin1973a,Bausch1976a,Janssen1976a,DeDominicis1978a,Zinnjustin2002a},
\begin{align}
 S[h,\tilde{h}] & = \int_\tx \tilde{h}\left(\partial_t h - \frac{1}{2} \left[\vec{\nabla}h\right]^2 - \nabla^2 h\right) \nonumber \\
 & \quad - \int_{t,\vec{x}_1,\vec{x}_2} \tilde{h}(t,\vec{x}_1)  \, R_\xi(\vec{x}_1-\vec{x}_2) \, \tilde{h}(t,\vec{x}_2) \, ,
 \label{eq:action}
\end{align}
which depends on the height field $h$ as well as the usual `response' field $\tilde{h}$. The non-local term in \Eq{eq:action}  arises because of the presence of the correlated noise ($\xi>0$). Terms related to initial conditions are neglected since we focus exclusively on the stationary state.

The \KPZ action \Eq{eq:action} possesses several symmetries. Apart from space-time translation and space rotation invariance, $S[h,\tilde{h}]$ is invariant under the following infinitesimal (terms of order $v^2$ and higher are neglected) field transformations:
\begin{subequations}
 \begin{alignat}{2}
& \left\{ \begin{array}{l} h'(\tx) = h(t,\vec{x}) + c \\
  \tilde{h}'(\tx) = \tilde{h}(t,\vec{x}) \end{array} \right. \, , \label{eq:shift} \\
& \left\{ \begin{array}{l} h'(\tx) = h(t,\vec{x}+ \vec{v} t) + \vec{v}\cdot \vec{x}\\
  \tilde{h}'(\tx) = \tilde{h}(t,\vec{x}+ \vec{v}t) \end{array} \right. \, . \label{eq:galilei}
\end{alignat}
\label{eq:sym_shft_galilei}%
\end{subequations}
For a \OneD interface and uncorrelated noise $\xi = 0$, the additional discrete transformation \cite{Canet2005b}
\begin{align}
& \left\{ \begin{array}{l} h'(\tx) = - h(-t,\vec{x})\\
  \tilde{h}'(\tx) = \tilde{h}(-t,\vec{x}) + \nabla^2 h(-t,\vec{x})\end{array} \right. \, ,
  \label{eq:time_reversal}
\end{align}
is also a symmetry. These transformations encode vertical shifts of the interface \eq{eq:shift}, Galilean boosts \eq{eq:galilei} and  time-reversal \eq{eq:time_reversal}. Moreover, the Galilean and shift symmetries can be gauged in time, considering $c(t)$ and $\vec{v}(t)$ as infinitesimal time-dependent quantities. The \KPZ action $S[h,\tilde{h}]$ is no longer invariant under the gauged transformations, but its change $S[h,\tilde{h}]-S[h',\tilde{h}']$ is linear in the fields. This provides generalised Ward identities with a stronger content than in the standard (non-gauged) case \cite{Lebdev1994,canet_2011_PhysRevE84_061128}. These symmetries play an important role in devising an accurate approximation scheme in the \FRG framework, see \sref{section-approximation}. We emphasise that the correlated noise explicitly breaks the time-reversal symmetry \eq{eq:time_reversal}. Similarly, a temporal correlation of the noise $\delta(t-t') \rightarrow R_{\xi_\tau}(t-t')$ would break the Galilean symmetry \eq{eq:galilei} \cite{Medina1989a,katzav2004,Fedorenko2007,Strack2014a,Song2016b}.

\subsection{Non-Perturbative Functional Renormalization Group}
\label{section-FRG}

The \FRG is a non-perturbative incarnation of the \RG (see \cite{Bagnuls:2000ae,Berges:2000ew,Polonyi:2001se,Delamotte:2007pf,Kopietz10} and references therein for  reviews, and in particular \cite{canet2011b,Berges2012a} for applications of the \FRG to non-equilibrium systems). It relies on Wilson's view of the \RG \cite{kadanoff1966a,wilson1971renormalizationI,wilson1971renormalizationII}, and consists in constructing a scale-dependent effective action, ${\Gamma_k[\varphi,\tilde{\varphi}]}$, where small spatial scales are integrated out. That is schematically
\begin{align}
 \text{e}^{-\Gamma_k[\varphi,\tilde{\varphi}]} = \int D\underset{p>k}{[h,\I\tilde{h}]} \text{e}^{-S[h,\tilde{h}]} \, , 
\end{align}
where Fourier modes with $p\leq k$ are frozen. $k$ is the momentum scale that separates small- (${p>k}$) and large-scale (${p<k}$) spatial  fluctuations, and ${\varphi \equiv \langle h \rangle}$ and ${\tilde \varphi \equiv \langle \tilde h \rangle}$ are the expectation values of the fields. In practice, the coarse-graining is achieved in a smooth way. To this end, a cut-off, or regulator, term\textsuperscript{\ref{ft:footnote_fourier}}
\begin{align}
 \Delta S_k[h,\tilde{h}] = \frac{1}{2} \int_\op h_i(\op) \mathcal{R}_{k,ij}(\vec{p}) h_j(\mop) \, ,
\end{align}
is added to the action, \Eq{eq:action}. $h_i$ for ${i = 1,2}$ label the  field $h$ and response field $\tilde{h}$ respectively, and repeated indices are summed over. The cut-off matrix $\mathcal{R}_{k}$ provides a momentum-dependent mass term to the theory. Its elements  are required to be of order $k$ (or higher) for ${p\lesssim k}$ and to vanish for $p\gtrsim k$. The regulator $\mathcal{R}_{k}$ must also vanish when the \RG scale $k$ is sent to $0$. Apart from these constraints, it can be chosen freely and will be specified in \Eq{eq:cut_off_matrix} below.  The flowing effective action ${\Gamma_k[\varphi,\tilde{\varphi}]}$  is defined (up to the additive $\Delta S_k$ term) as the Legendre transform of the logarithm of the generating functional of the coarse-grained theory ${W_k[J,\tilde{J}] = \ln(Z_k[J,\tilde{J}])}$:
\begin{align}
 & Z_k[J,\tilde{J}] = \int D[h,\I\tilde{h}] \, \text{e}^{-S-\Delta S_k + \int_\tx \left(h J + \tilde{h} \tilde{J}\right)} \, , \nonumber \\
 & \Gamma_k[\varphi,\tilde{\varphi}] + \Delta S_k[\varphi,\tilde{\varphi}] \coloneqq \nonumber \\
 & \qquad \underset{J,\tilde{J}}{\text{Sup}}\left\{ \int_{\tx} \left(\varphi J + \tilde{\varphi} \tilde{J} \right)-W_k[J,\tilde{J}] \right\} \, .
 \label{eq:gammak_def}
\end{align}
The addition of the $\Delta S_k$ term ensures that $\Gamma_k$ interpolates (as the \RG scale  $k$ is decreased from the \UV scale\footnote{$\Lambda$ can be interpreted as the inverse lattice length of a discrete system.} $\Lambda$ to zero) in between microscopic and macroscopic physics. The bare action \Eq{eq:action} is recovered in the limit of large $k=\Lambda$ and the full 1-particle irreducible effective action (which is analogous to the Gibbs free energy in thermodynamics) is obtained in the limit $k\to 0$ where the cut-off is removed,
\begin{align}
 \Gamma_{k\rightarrow \Lambda}[\varphi,\tilde{\varphi}] = S[\varphi,\tilde{\varphi}] \, , && \Gamma_{k\rightarrow 0}[\varphi,\tilde{\varphi}] = \Gamma[\varphi,\tilde{\varphi}] \, .
 \label{eq:gammak_init}
\end{align}

The definition of the Legendre transformation \Eq{eq:gammak_def} relates the sources $J,\tilde{J}$ to the fields $\varphi,\tilde{\varphi}$ through
\begin{subequations}
 \begin{alignat}{2}
  & \frac{\delta W_k[J,\tilde{J}]}{\delta J_i(\tx)} && = \varphi_i(\tx) \, , \label{eq:htoj}\\
 & \frac{\delta \Gamma_k[\varphi,\tilde{\varphi}]}{\delta \varphi_i(\tx)} &&= J_i(\tx) - \frac{\delta \Delta S_k[\varphi,\tilde{\varphi}]}{\delta \varphi_i(\tx)}\, . \label{eq:jtoh}
 \end{alignat}
 \label{eq:sources}%
\end{subequations}
The (connected) two-point correlation and response functions at scale $k$,
\begin{align}
 & G_{k,ij}(t-t',\vec{x}-\vec{x}')  = \moy{h_i(\tx) h_j(\txp)}_c \, ,
 \label{eq:corr}
\end{align}
are given by the operator inverse of the second field derivative of $\Gamma_k+\Delta S_k$,
\begin{align}
 G_k = \frac{1}{\Gamma_k^{(2)}+\mathcal{R}_k} \, ,
 \label{eq:green}
\end{align}
with the notation
\begin{align}
\Gamma_{k,ij}^{(2)}(t-t',\vec{x}-\vec{x}') = \frac{\delta^2 \Gamma_k[\varphi,\tilde{\varphi}]}{\delta \varphi_i(t,\vec{x}) \delta \varphi_j(\txp)} \, .
\end{align}
Because of space-time translational invariance, the second field derivative of $\Gamma_k$ is diagonal in momentum space. This implies that the Fourier transform of $C(\taur)$ [\Eq{eq:def_c}] is simply obtained from the matrix inverse of  ${\Gamma_k^{(2)}(\op)}$ in the limit $k\rightarrow 0$ as
\begin{align}
 C(\op) = \left[\Gamma_{k,ij}^{(2)}(\op) \right]^{-1}_{11} \, .
 \label{eq:comp_c}
\end{align}

In principle, the choice of the regulator matrix does not affect the end results. However, in practice, approximations introduce a spurious dependence on $\mathcal{R}_{k}$. Since symmetries provide strong constraints on the space of solutions of the \RG flow equations, an important requirement is that $\Delta S_k[h,\tilde{h}]$ preserves the symmetries of the theory. We choose
\begin{align}
 \mathcal{R}_{k}(\vec{p}) = \frac{\alpha}{\text{e}^{\frac{p^2}{k^2}}-1}\left(\begin{array}{cc} 0 & \nu(k) p^2 \\ \nu(k) p^2 & -2 D(k) \end{array}\right) \, ,
 \label{eq:cut_off_matrix}
\end{align}
where $\nu(k)$ and $D(k)$ are two coefficients that depend on the \RG scale $k$. They will be defined in the next section, \Eq{eq:dknuk}. The coefficient  $\alpha$ is a free parameter that can be tuned to minimise the errors at a given order of approximation \cite{Stevenson1981,Canet:2002gs} (see \aref{sec-pms}). Note that $\mathcal{R}_{k,ij}$ has the same tensor structure as the bare propagator (second field derivative of $S[h,\tilde{h}]$) and does not depend on frequency. This ensures that  the coarse-grained theory is causal, and that the flow preserves the Galilean and shift symmetries \Eqs{eq:sym_shft_galilei}, and also, when $d=1$ and $\nu(k) = D(k)$, the time-reversal symmetry \eq{eq:time_reversal} \cite{canet_2011_PhysRevE84_061128}.

The evolution of the effective action $\Gamma_k$ with the \RG scale is given by an exact equation \cite{Wetterich:1992yh},
\begin{align}
 k\partial_k \Gamma_k[\varphi,\tilde{\varphi}] = \frac{1}{2} \text{Tr}\left[\frac{k\partial_k \mathcal{R}_k}{\Gamma_k^{(2)}[\varphi,\tilde{\varphi}]+\mathcal{R}_k} \right] \, ,
 \label{eq:wetterich}
\end{align}
with its initial condition corresponding to the bare KPZ action \Eq{eq:action}, as stated by \Eq{eq:gammak_init}. The trace operation on the  right-hand side stands for the usual trace over field and space-time indices,
\begin{align}
 \text{Tr}\left[A\right] = \sum_i \int_\tx A_{ii}(\tx) \, .
\end{align}

{\Eq{eq:wetterich} provides a scheme to include fluctuations gradually starting with the small-scale fluctuations and reaching  the thermodynamic limit ($k\rightarrow 0$). At intermediate values of $k$, $\Gamma_k$ plays different roles for large and small momenta (compared to $k$). Derivatives of $\Gamma_k$ with respect to fields with small momenta ($p\ll k$) yield the kinetic term and the vertices of an effective action that can be used [instead of the original bare action \Eq{eq:action}] to compute large-scale correlation functions. On the other hand, when the momenta are large ($p\gg k$), the derivatives of $\Gamma_k$ quickly lose their dependence on $k$ and saturate to their physical values (as $k$ is lowered further). In this regime correlation functions are computed directly (with no further functional integration) by the procedure outlined above [\Eqs{eq:green} and \eq{eq:comp_c}].

Note that the \UV cut-off scale should (in principle) be taken to infinity to describe the continuous \KPZ equation. In practice, it is sufficient to choose it to be much larger than all the momentum scales that are resolved. In particular, one must have $\Lambda \gg 1/\xi$ to probe the structure of the microscopic noise. Conversely, when $\xi \ll 1/\Lambda$,  the $\delta$-correlated case is effectively described.

\subsection{Approximation scheme}
\label{section-approximation}

\Eq{eq:wetterich} combined with \Eqs{eq:gammak_init} provides a differential equation and an initial condition (at $k\rightarrow \Lambda$). In principle, $\Gamma_k[\varphi,\tilde{\varphi}]$ can hence be determined for all values of $k$. However, \Eq{eq:wetterich} is a functional partial differential equation that cannot be solved exactly. It couples derivatives of $\Gamma_k$ of order $n$ to derivatives of order $(n+1)$ and $(n+2)$ and generates an infinite hierarchy of equations relating all the correlation functions of the problem.

Here we use a very successful approximation scheme developed for the \KPZ equation with $\delta$-correlated noise ($\xi=0$) \cite{kloss_canet_wschebor_2012_PhysRevE86_051124}, that is inspired by the \BMW approximation \cite{Blaizot:2005xy,Benitez2012}, but rendered compatible with the constraining symmetries of the \KPZ action (see \eg \cite{canet_2011_PhysRevE84_061128} for a detailed description). A practical way to implement this scheme is to construct an ansatz for the flowing effective action $\Gamma_k$, which automatically preserves the gauged Galilean symmetry, by using explicitly Galilean invariant building blocks. In particular, this involves the covariant time derivative
\begin{align}
 \widetilde{D}_t = \partial_t - \vec{\nabla}\varphi \cdot \vec{\nabla} \, ,
 \label{eq:covariant_derivative}
\end{align}
which preserves the invariance under Galilean transformation. When it is truncated to \SO in the response field $\tilde{\varphi}$, the ansatz obtained with this procedure is
\begin{align}
 & \Gamma_k[\varphi,\tilde{\varphi}] = \int_{\tx}\Big[  \tilde{\varphi} f^{\lambda}_k \left(\partial_t \varphi - \frac{1}{2} \left[\vec{\nabla}\varphi\right]^2 \right) \nonumber \\
& \quad - \frac{1}{2} \left( \vec{\nabla}^2 \varphi f^{\nu}_k \tilde{\varphi} + \tilde{\varphi} f^{\nu}_k \vec{\nabla}^2 \varphi \right) +  \tilde{\varphi} f^{D}_k \tilde{\varphi} \Big]\, .
\label{eq:ansatz}
\end{align}
$f_k^X$ (with $X \in \{ \lambda,\nu,D \}$) are analytic functions of $\widetilde{D}_t$ and $\vec{\nabla}$, which depend on the \RG scale. They can be interpreted as an effective non-linearity, dissipation and noise respectively. The bare action \Eq{eq:action}, is recovered when
\begin{align}
 f_\Lambda^\nu(\omega,\vec{p}) = 1 \, , && f_\Lambda^\lambda(\omega,\vec{p}) = 1 \, , && f_\Lambda^D(\omega,\vec{p}) = R_\xi(p) \, .
 \label{eq:initial}
\end{align}
When they are evaluated at a uniform and stationary configuration ($\varphi = \text{const.}$ and $\tilde{\varphi}=0$), the derivatives of $\Gamma_k[\varphi,\tilde{\varphi}]$ in Fourier space become expressions depending on $f_k^X(\omega,\vec{p})$, that is the operators $\widetilde{D}_t$ and $\vec\nabla$ in $f_k^X$ are replaced by $i\omega$ and $-i \vec{p}$ respectively. Note that the ansatz \Eq{eq:ansatz} contains arbitrary powers of $\varphi$ through the functional dependence of $f_k^X$ on the covariant time derivative $\widetilde{D}_t$. 

There are additional constraints on $f_k^X$ stemming from the other symmetries. The gauged shift symmetry imposes
\begin{align}
 f_k^\lambda(\omega,\vec{0}) = 1 \, .
 \label{eq:gauge_shift_ward}
\end{align}
For $\xi=0$ and $d=1$, the time-reversal symmetry, \Eq{eq:time_reversal}, leads to
\begin{align}
f_k^D = f_k^\nu\, , && f_k^\lambda = 1 \, ,
 \label{eq:fdt}
\end{align}
such that there is only one independent function left in this case.

In the following, we focus on the \OneD case (vector symbols are hence dropped). With the ansatz \Eq{eq:ansatz}, the inverse propagator evaluated at uniform and stationary configuration reads
\begin{align}
 & \Gamma_k^{(2)}(\omega,p) = \nonumber \\
 & \left(\begin{array}{cc} 0 & i\omega f_k^{\lambda}(\omega,p) + p^2 f_k^{\nu}(\omega,p) \\ - i\omega f_k^{\lambda}(\omega,p) + p^2 f_k^{\nu}(\omega,p) & -2 f_k^D(\omega,p) \end{array} \right) \, .
\label{eq:inverse_propagator}
 \end{align}
This is the most  general form for $\Gamma_k^{(2)}$ compatible with the symmetry constraints, and endowed with an arbitrary dependence on $\omega$ and $p$. On the other hand, higher-order vertices $\Gamma_k^{(n>2)}$ are  approximated.

When the ansatz for ${\Gamma_k}$ \Eq{eq:ansatz}, is inserted into its exact evolution equation \Eq{eq:wetterich}, the \RG flow can be projected onto the flows of $f_k^X$. This provides three partial differential equations that can be solved numerically,
\begin{align}
 & k\partial_k f_k^X(\omega,p) = I_k^X(\omega,p)  = \int_{f,q} J_k^X\left(\omega,p,f,q\right) \, ,
 \label{eq:dimensionfull_flow}
\end{align}
where $I_k^X$ are non-linear integral expressions depending on $f_k^X$, which can be found in \cite{kloss_canet_wschebor_2012_PhysRevE86_051124}. They are obtained by taking appropriate field derivatives of the right-hand side of the exact flow equation \Eq{eq:wetterich}, and replacing $\Gamma_k$ by its ansatz. The trace in \Eq{eq:wetterich} produces integral equations with non-linear kernels $J_k^X$.

The \SO approximation is further simplified to the so-called \NLO approximation, introduced in \cite{kloss_canet_wschebor_2012_PhysRevE86_051124}. It consists in  partially truncating the frequency dependence of $f_k^\nu$ and $f_k^D$ by setting $f_k^X(\Omega,Q) = f_k^X(0,Q)$ in $J_k^X(\omega,p,f,q)$ on the right-hand side of \Eq{eq:dimensionfull_flow} for any arguments ${Q=q, |p\pm q|}$, and similarly for $\Omega$. This simplification drastically reduces the computational cost of solving \Eqs{eq:dimensionfull_flow} while still yielding reliable results (see \cite{kloss_canet_wschebor_2012_PhysRevE86_051124}). Moreover, in the presence of microscopic noise correlations (${\xi\geq 0}$), we consider two independent flowing functions $f_k^\nu$ and $f_k^D$ since \Eqs{eq:fdt} are not satisfied for $\xi \neq 0$. However, we  still impose $f_k^\lambda = 1$, following \cite{kloss_canet_wschebor_2012_PhysRevE86_051124} \footnote{\label{ft:footnote_flambda}In fact, within the NLO approximation, some higher-order vertices, which are involved in the Ward identity  related to time-reversal symmetry yielding $f_k^\lambda=1$, are neglected. It implies that this identity cannot be restored exactly even when the flow leads to a  time-reversal symmetric fixed-point. Hence, setting $f_k^\lambda=1$ prevents $f_k^\lambda$ from  acquiring a non-trivial flow which would induce a residual small breaking of time-reversal symmetry when the latter should be realised in the \IR. However, the related error is  small. It was checked in \cite{kloss_canet_wschebor_2012_PhysRevE86_051124}  that the obtained exponents and dimensionless ratio differ only weakly with or without this approximation. Note that, on the other hand,  no constraints are imposed on $f_k^\nu$ and $f_k^D$, which are free to be different, such that the complete \RG flow is not constrained to be time-reversal symmetric.}.

To summarise, our approximation scheme consists of three main points:
\begin{itemize}
 \item  $\Gamma_k$ is expanded in powers of $\tilde{\varphi}$ and only terms up to second order are retained. This is the \SO approximation level, which is essential to produce a manageable set of flow equations. It amounts to assuming that the fluctuations of the response field $\tilde{h}$, are Gaussian. Note that this truncation is not applied to $h$ since $\Gamma_k$ contains arbitrarily high powers of $\varphi$. The \SO was shown in \cite{canet_2011_PhysRevE84_061128} to reproduce the exact results for the \OneD scaling function when $\xi=0$ \cite{Prahofer2004} with an unprecedented accuracy.
 \item The frequency dependence of $f_k^X$ is neglected in the kernel $J_k^X$, on the right-hand side of the flow equation (\ref{eq:dimensionfull_flow}). This is the \NLO approximation level. This approximation can be assessed by comparing its outcome to the results of the \SO approximation alone. This was done for $d=1$ in \cite{kloss_canet_wschebor_2012_PhysRevE86_051124} and only small differences were observed. This approximation greatly speeds up the numerical solution of the flow equations since it enables the analytical integration over the internal frequencies $f$, in \Eq{eq:dimensionfull_flow}. Note that the bare frequency content is preserved so that $f_k^X$ still develops a non-trivial frequency dependence. For $d=2$ and $3$ and $\xi=0$, the exponents as well as universal dimensionless ratios computed in \cite{kloss_canet_wschebor_2012_PhysRevE86_051124} at \NLO are in close agreement with the outcome of numerical simulations \cite{HalpinHealy2013a,HalpinHealy2013b}.
 \item  $f_k^\lambda$ is set to one. This approximation is specific to \OneD because it is related to the time-reversal symmetry \Eq{eq:time_reversal}.  It ensures that, if the long-distance physics is described by a time-reversal symmetric \IR fixed point, then the latter is recovered exactly, whereas if $f_k^\lambda$ starts flowing, it induces a small error on the fixed point properties. See footnote \ref{ft:footnote_flambda} for details.
\end{itemize}

\section{Universal and non-universal features of the two-point correlation function}
\label{section-NPRG-universal-vs-nonuniversal}

\subsection{Fixed point}
\label{subsection-RG-flow}

In order to find \RG fixed points and study scale invariance, it is convenient to recast \Eqs{eq:dimensionfull_flow} in a dimensionless form. To this end we define
\begin{align}
 \nu(k) = f_k^{\nu}(0,0)\, , && D(k) = f_k^{D}(0,0)\, ,
 \label{eq:dknuk}
\end{align}
and introduce the rescaled variables\footnote{One can check in \Eq{eq:inverse_propagator} that the introduction of the coefficients $\nu(k)$ and $D(k)$ in the regulator matrix \Eq{eq:cut_off_matrix} ensures that the cut-off term scales (with $k$) as the rest of the kinetic term of $\Gamma_k$. In this way, none overwhelms the other (as the cut-off scale decreases) and the cut-off matrix stays effective for all $k$ in the rescaled units.},
\begingroup
\allowdisplaybreaks
\begin{align}
 & \hat{p} = \frac{p}{k} \, , \quad \hat{\omega} = \frac{\omega}{\nu(k) k^2} \, , \nonumber \\
 & \hat{f}_k^\nu(\hat{\omega},\hat{p}) = \frac{f_k^\nu(k^2\nu(k) \hat{\omega},k \hat{p})}{\nu(k)} \, ,   \nonumber \\ &\hat{f}_k^D(\hat{\omega},\hat{p}) = \frac{f_k^D(k^2\nu(k) \hat{\omega},k \hat{p})}{D(k)} \, ,\nonumber \\
& \hat{h}(\hat\omega,\hat p) =\sqrt{\frac{k^7\nu(k)^3}{D(k)}} h(\omega,p) \, ,   \nonumber \\
&\hat{\tilde{h}}(\hat\omega,\hat p) = \sqrt{k^3 D(k) \nu(k)} \tilde{h}(\omega,p)\, .
\label{eq:rescalings}
\end{align}
\endgroup
The flows of $\nu(k)$ and $D(k)$ define the two running anomalous dimensions,
\begin{align}
 \eta_\nu(k) = - \frac{k\partial_k \nu(k)}{\nu(k)} \, , &&\eta_D(k) = - \frac{k\partial_k D(k)}{D(k)} \, .
 \label{eq:anomalous}
\end{align}
In terms of the rescaled quantities, the ansatz for $\Gamma_k$ bares the same form as its original definition \Eq{eq:ansatz} but for the term in $[\hat \nabla \hat \varphi]^2$ that is multiplied by ${\sqrt{\hat g_k}}$ [with ${\hat g_k = D(k)/(k \nu(k)^3)}$] and for the covariant time derivative that is changed to  $\widetilde{D}_{\hat t} = \partial_{\hat t} - \sqrt{\hat g_k} \, {\hat \nabla}\hat{\varphi} \cdot {\hat\nabla}$. Since $\lambda$ is not renormalized due to Galilean invariance, the flow of $\hat{g}_k$ is only dimensional and reads
\begin{align}
 k\partial_k \hat g_k = \hat g_k \, [3\eta_\nu(k)-1 -\eta_D(k)] \ .
 \label{eq:dkgk}
\end{align}
In terms of the rescaled variables, $\Gamma_k$ depends on the cut-off scale only implicitly through $\hat{f}_k^X$. This enables the emergence of solutions of the flow equations, \Eqs{eq:dimensionfull_flow} [see \Eqs{eq:flow} for the rescaled equations] where $\hat{f}_k^X$, $\eta_X(k)$ (with $X=\nu$, $D$) and $\hat{g}_k$ do not depend on $k$. These are fixed points of the \RG flow and describe the universal properties of the system.

At such a fixed point, the running anomalous dimensions tend to constant values $\eta_D(k) \to \eta_D^* \coloneqq \eta_D$ and $\eta_\nu(k) \to  \eta_\nu^*\coloneqq \eta_\nu$. This implies that $\nu(k) = \nu_\xi k^{-\eta_\nu}$ and $D(k) = D_\xi k^{-\eta_D}$ behave as power laws, where $\nu_\xi$ and $D_\xi$ are non-universal constants that cannot be determined from the fixed point alone but can be extracted from the full solution of the flow,
\begin{alignat}{2}
& \nu(k) = \text{e}^{\int_{k}^{\Lambda} \frac{\eta_\nu(k')}{k'} \text{d}k'} && \xrightarrow[k \rightarrow 0]{} \nu_\xi \, k^{-\eta_\nu} \, , \nonumber \\
& D(k) = \text{e}^{\int_{k}^{\Lambda} \frac{\eta_D(k')}{k'} \text{d}k'} && \xrightarrow[k \rightarrow 0]{} D_\xi \, k^{-\eta_D} \, .
\label{eq:expoflow}
\end{alignat}
The dimensionful two-point correlation function [defined in \Eq{eq:def_c} and computed from \Eq{eq:comp_c}] can be expressed in terms of rescaled quantities as\textsuperscript{\ref{ft:footnote_fourier}}
\begin{align}
  C(\omega,p) &= 
   k^{-4} \frac{D(k)}{\nu(k)^2} \hat C_k\left(\frac{\omega}{k^2\nu(k)},\frac{{p}}{k}\right) \nonumber\\
 & = \frac{D_\xi}{\nu_\xi^2} \, k^{2\eta_\nu-\eta_D-4} \, \hat C_*\left(\frac{\omega}{\nu_\xi k^{2-\eta_\nu}},\frac{{p}}{k}\right) \, ,
 \label{eq:rescaled_two_point}
\end{align}
where ${\hat{C}_k(\hat{\omega},\hat{p})}$ depends on $k$ only through ${\hat{f}_k^X}$. The second equality holds for $k$ small enough such that the fixed point has been reached. Then $k$ is a free parameter and can be chosen to be $k=p$. Identifying with the correlator scaling form \Eq{eq:chi_and_z}, in Fourier space, yields
\begin{align}
 \chi = (\eta_D+1-\eta_\nu)/2 \, , && z = 2-\eta_\nu \, .
 \label{eq:eta2chi}
\end{align}
Furthermore, \Eq{eq:dkgk} enforces that $z+\chi=2$ at any non-Gaussian fixed point $\hat g_*\neq 0$. Note that at such a fixed point $\eta_X$ (and therefore $\chi$ and $z$) could be functions of $\xi$, but this is not the case, as shown in the following.

The flow equation (\ref{eq:dimensionfull_flow}) becomes, in terms of the rescaled quantities,
\begin{align}
 k\partial_k \hat{f}_k^X = \eta_X(k) \hat{f}_k^X + \left[2-\eta_\nu(k)\right]  \, \hat{\omega} \partial_{\hat{\omega}} \hat{f}_k^X+ \hat{p} \partial_{\hat{p}} \hat{f}_k^X + \hat{I}^X \, .
 \label{eq:flow}
\end{align}
$\hat{I}^X(\hat{\omega},\hat{p})$ is obtained from $I_k^X$ \Eq{eq:dimensionfull_flow}, by switching to the rescaled variables and dividing by $\nu(k)$ or $D(k)$ accordingly. The equations for $\eta_X(k)$ are deduced by evaluating \Eqs{eq:flow} at zero momentum and frequency and fixing $\hat{f}_k^X(0,0)=1$ consistently with \Eqs{eq:dknuk} and \eq{eq:rescalings}. Their explicit expressions are given in \cite{kloss_canet_wschebor_2012_PhysRevE86_051124}.

The initial condition \Eq{eq:initial} of the flow is specified at a large but finite \UV scale $\Lambda$, and reads for the dimensionless quantities as
\begin{align}
  \hat{f}_\Lambda^\nu(\hat{\omega},\hat{p}) = 1 \, , && \hat{f}_\Lambda^D(\hat{\omega},\hat{p}) 
  = R_{\hat{\xi}}(\hat{p}) \, , && \hat g_\Lambda = \frac{1}{\Lambda} \, ,
  \label{eq:initial_rescaled}
\end{align}
with $\hat{\xi} = \Lambda \xi$. In the unit system defined by \Eq{eq:kpz_du}, $\Lambda$ (as well as $k$) is dimensionless. We have chosen $\Lambda = 200$. Note that the actual value of $\Lambda$ never enters the dimensionless \RG flow. It is only necessary to commit to an actual value when dimensionful quantities are computed. Moreover, even though it seems that an additional parameter is needed [$g$ is set to one at the beginning, \Eq{eq:kpz_du}] to specify the initial conditions in the rescaled variables, only the specific combination $\hat{\xi}/\hat g_\Lambda \rightarrow \xi$ is physically observable (does not depend on $\Lambda$) in the continuum limit $\Lambda \rightarrow \infty$. We choose $\hat{g}_\Lambda = 1/\Lambda$ and $\hat{\xi} = \Lambda \xi$ which is consistent with the choice of units \Eq{eq:kpz_du}. This implies that the \RG flow starts infinitesimally close to a Gaussian fixed point ($\hat{g}_{\Lambda\rightarrow \infty}=0$) and evolves towards the non-linear \KPZ physics on large scales (${\hat{g}_{k \to 0} \neq 0}$).

We have solved \Eqs{eq:flow} and (\ref{eq:dkgk}) numerically for different values of $\hat{\xi}=\Lambda\xi$. We found two remarkable properties:
\begin{itemize}
 \item Although the microscopic action is not invariant under the time-reversal symmetry, \eq{eq:time_reversal} (\ie ${\hat{f}_\Lambda^D \neq \hat{f}_\Lambda^\nu}$), both functions tend to each other as ${k\rightarrow 0 }$. A fixed point is reached, with $\hat{f}_{*}^D = \hat{f}_{*}^\nu$. This means that the time-reversal symmetry is emergent. Even if it is explicitly broken by the microscopic theory (${k\rightarrow \Lambda}$), it is realised on large scales (${k\rightarrow 0}$).
 \item The flow reaches the same fixed point for all values of $\hat{\xi}$. The theories with $\xi > 0 $ are in the basin of attraction of the standard ($\xi=0$) \KPZ fixed point \cite{Kardar1986a,Sasamoto2010a,Calabrese2011,Corwin2012a}. This implies that the large-scale physics is universal, independent of the details of the microscopic noise, and governed by the ${\xi=0}$ fixed point. In particular, the exponents of the scaling regime are the ones of the standard \OneD \KPZ equation with an uncorrelated noise, that is $\chi = 1/2$ and $z=3/2$.
\end{itemize}
This behaviour is illustrated in \fref{fig:rg_flow} which displays $\hat{f}_k^D(0,\hat p)$ and $\hat{f}_k^\nu(0,\hat p)$ for different values of $\hat{\xi}$, and at successive \RG `times' $s=\ln(k/\Lambda)$. Starting from initial conditions \Eq{eq:initial_rescaled} with different $\hat{\xi}$ at $s=0$ (upper set of plots), the functions evolve under the \RG flow, until they coincide at $s \leq -6$ (third set of plots), and then they deform to reach their fixed point shape, represented in the fourth set of plots.
\begin{figure}[ht]
\includegraphics[width=\columnwidth]{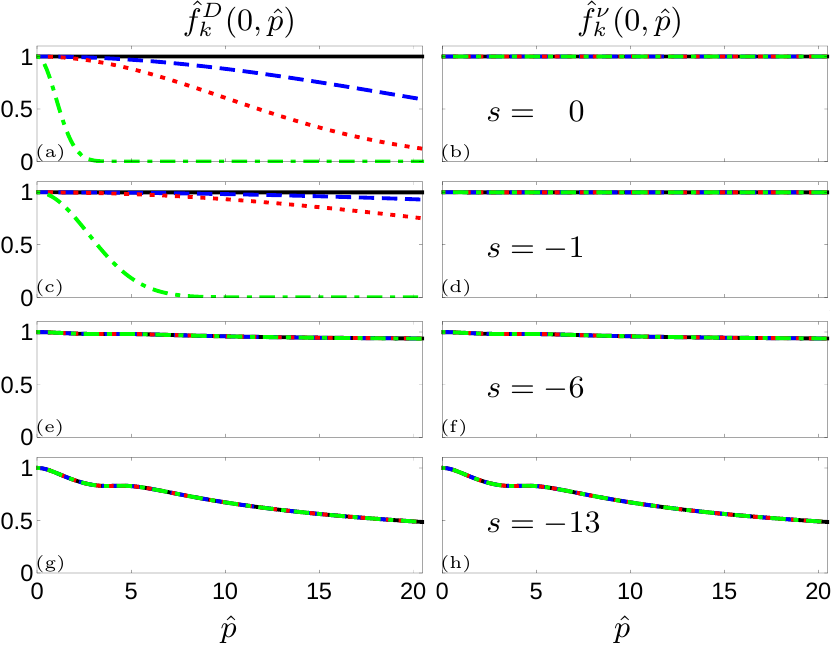}
\caption{(Colour online) \RG flow of the effective noise correlation $\hat{f}_k^D(0,\hat{p})$ ((a), (c), (e) and (g)), and dissipation $\hat{f}_k^\nu(0,\hat{p})$ ((b), (d), (f) and (h)) - Different values of the rescaled correlation length are shown: $\hat{\xi}=0$ (black, solid), $\hat{\xi}=0.05$ (blue, dashed), $\hat{\xi} = 0.1$ (red, dotted) and $\hat{\xi}=1$ (green, dot-dashed). The two functions are plotted as functions of the rescaled momentum $\hat{p}$, for $\hat{\omega}=0$ and for four different values of the \RG  `time', $s=\ln(k/\Lambda)$, as the cut-off scale is lowered $s = 0$ ((a) and (b)), $s=-1$ ((c) and (d)), $s=-6$ ((e) and (f)) and $s=-13$ ((g) and (h)). The axes of all the plots are scaled in the same way. One sees that the time-reversal symmetry ($\hat{f}_k^D=\hat{f}_k^\nu$), which is broken at $s=0$, has emerged at $s \leq -6$ and that the same fixed point is reached for all values of $\hat{\xi}$ when $s\leq -13$.}
\label{fig:rg_flow}
\end{figure}

\subsection{Scaling of the two-point correlator}
\label{subsection-two-point-correlator}

The fact that the \RG flow for finite $\xi$ leads to the same \IR fixed point as for $\xi=0$ implies that the two-point correlation function endows on large scales a scaling form with the same universal scaling function, denoted $\hat F$, as the $\xi=0$ case. Let us first emphasise that the fixed point is fully attractive: Our numerical analysis shows that it is reached for any initial condition without the need to fine-tune any parameter (no unstable direction). This means that the \KPZ dynamics leads to generic scale invariance in \OneD, as expected physically.
    
The existence of a scaling form for the correlation function ${C(\omega,p)}$ was shown in \cite{canet_2011_PhysRevE84_061128} for $\xi=0$. It relies on both the existence of the fixed point, that is $k\partial_k \hat f^X_k=0$ in \Eq{eq:flow}, and the decoupling property, that is, $\hat I^X(\hat \omega,\hat p)\to 0$  when  $\hat p \gg 1$ or/and $\hat \omega \sim \hat p^{3/2} \gg 1$. This property induces the flow to essentially stop for the large $p$ or/and $\omega$ sectors of $\Gamma_k$, which hence decouple from the other sectors. When both these conditions are satisfied, the general solution of the remaining homogeneous equation \Eq{eq:flow} for large $\hat p$ is a scaling form
\begin{equation}
 \hat f^X_*(\hat \omega,\hat p) = \hat p^{-\eta_X} \zeta^X(\hat \omega /\hat p^{3/2})\,.
 \label{eq:scaling}
 \end{equation}
The scaling form emerges in $\hat{f}_k(\hat{\omega},\hat{p})$ only at intermediate values of $\hat{p}$ because the rescaled functions do not tend to their fixed point value uniformly. In fact, for $\xi\neq 0$, the latter is reached when $\hat{p} \ll 1/(k\xi)$ (or $\hat{p} \ll \Lambda/k$ when $\xi \ll 1/\Lambda$, see the discussion at the end of \sref{section-FRG}). In the rescaled units, the non-universal features are gradually sent to larger and larger values of $\hat{p}$ as $k$ decreases. This implies that a scaling range emerges, where \Eq{eq:scaling} holds, for ${1 \ll \hat p \ll 1/(k\xi)}$, that is when the fixed point is reached and decoupling has occurred. Switching back to the dimensionful quantities and using \Eq{eq:expoflow}, it follows that
\begin{align}
 f_{k\rightarrow 0}^X(\omega,p) = X_\xi \, p^{-\eta_X} \zeta^X\left(\frac{\omega}{\nu_\xi p^{3/2}}\right) \, ,
\end{align}
for ${k\ll p \ll 1/\xi}$. In practice, the scaling function is extracted from the numerical solution as
\begin{align}
\zeta^X\left(\hat{x}\right) = \lim_{\hat{p}\rightarrow \infty} \hat{p}^{\eta_X} \hat{f}_*^X\left(\hat{x}\hat{p}^{3/2},\hat{p}\right) \, .
\label{eq:fixed_point_scaling}
\end{align}
We emphasise that the exponent $\eta_X$ and the scaling functions $\zeta_X$ are universal, whereas  ${X_\xi}$ and $\nu_\xi$ are non-universal and depend explicitly on the correlated microscopic noise.

The two-point correlation function can be determined from $\Gamma_k$ through \Eq{eq:comp_c} and the inverse of $\Gamma_k^{(2)}$, \Eq{eq:inverse_propagator},
\begin{align}
 C(\omega,p) = \lim_{k\rightarrow 0} \frac{2 f_k^D(\omega,p)}{\omega^2+\left[p^2 f_k^\nu(\omega,p)\right]^2}  \, .
 \label{eq:c_from_f}
\end{align}
One then deduces that, in the regime ${k \ll p \ll 1/\xi}$, the dimensionful correlation function also takes a scaling form
\begin{align}
 C(\omega,p) = \frac{D_\xi}{\nu_\xi^2}\,p^{-7/2} \, \hat F\left(\frac{\omega}{\nu_\xi \, p^{3/2}}\right) \, ,
 \label{eq:scaling_g}
\end{align}
where the scaling function $\hat F$ is the same for any $\xi$, and can be determined from the fixed point solution as 
\begin{align}
\hat F(\hat{x}) = \frac{2 \zeta^{D}(\hat{x})}{\hat{x}^2+\left[\zeta^{\nu}(\hat{x})\right]^2} \, .
 \label{eq:Gxi}
\end{align}
We have inserted into \Eq{eq:scaling_g} the values ${\eta_\nu = \eta_D = 1/2}$, known from \Eq{eq:eta2chi}. For different values of $\xi$, the scaling form of the correlation functions hence only differs by $\xi$-dependent non-universal amplitudes that can be extracted from the \RG flow.

This is confirmed by the numerical solution of the flow. We computed $C(\omega,p)$ for different values of $\xi$ (including $\xi=0$). To select the regime where scale invariance is expected, we introduce an auxiliary scale $K$, such that momenta and frequencies 
\begin{align}
 p > K \, , && \omega > K^{3/2} \, ,
\end{align}
are excluded. Then, the truncated correlation function ${C^K(\omega,p)}$ is multiplied by $p^{7/2}$ and recorded as a function of the scaling variable $x = \omega/p^{3/2}$. This provides the scaling function 
\begin{align}
 F_\xi^K(x) = p^{7/2} \,C^K(x p^{3/2},{p}) \, ,
 \label{eq:non_universal_scaling}
\end{align}
which is related to the universal scaling function \Eq{eq:Gxi} by normalisation factors,
\begin{align}
F_\xi{^K}(x) = D_\xi/\nu_\xi^2 \, \hat{F}(x/\nu{_\xi}) \, . 
\label{eq:scaling_xi}
\end{align}
For each value of $\xi$, a collapse is indeed observed for $K$ small enough. Furthermore, ${\hat{F}}$ is the same for all $\xi$. This is illustrated in \fref{fig:collapse}, see \aref{sec-appendix-extracting} for more details.
\begin{figure}[ht]
\includegraphics[width=0.95\columnwidth]{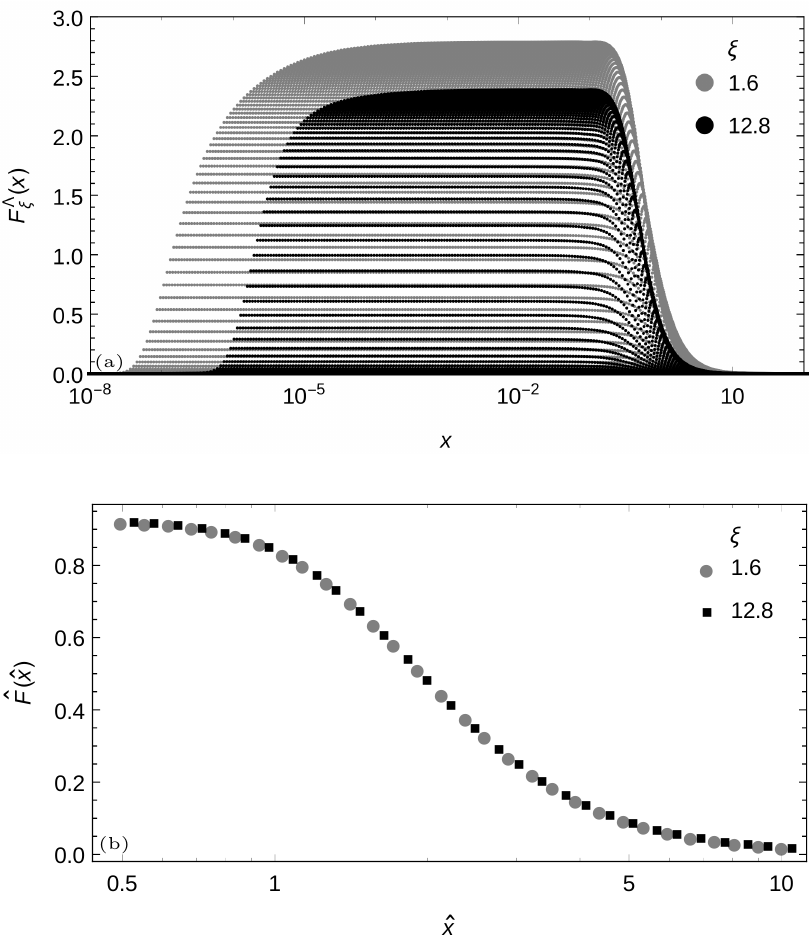}
\caption{Collapse of two-point correlation functions on large scales for two different values of the microscopic noise correlation length, $\xi = 1.6$ (grey) and $\xi = 12.8$ (black) - (a): The scaling function $F_\xi^\Lambda(x)$ [see \Eq{eq:non_universal_scaling}], is represented as a function of $x$ for all momenta and frequencies. This plot shows that different values of $p$ lead to different functions of $x$. There is no data collapse. (b): The universal scaling function $\hat{F}(\hat{x}) = \nu_\xi^2/D_\xi \, F_\xi^{0.001}(\nu_\xi \, \hat{x})$, is represented as a function of $\hat{x}$ for $p<K$ and $\omega<K^{3/2}$ with $K=0.001$. The normalisation factors $\nu_\xi$ and $D_\xi$ are obtained from the fitting procedure detailed in \aref{sec-appendix-extracting} and are consistent with the solution of the \RG flow equations, \Eqs{eq:expoflow}. It is clear from (b) that when $K=0.001$, enough \UV modes are excluded and $F_{1.6}^{0.001}$ and $F_{12.8}^{0.001}$ are equivalent (up to non-universal normalisation factors).}
\label{fig:collapse}
\end{figure}

\subsection{Non-universal correlations}
\label{sec:non_universal_correlations}

The \FRG is not restricted to the study of the universal properties of a system that emerge close to a fixed point. In this section we compute the (non-universal) kinetic energy spectrum in the stationary state for all momenta, and not only momenta in the scaling regime. We compare these results to the outcome of previous numerical simulations presented in \cite{agoritsas_2012_FHHtri-numerics}. Finally, we resolve the $\xi$-dependent crossover of the amplitude of the kinetic energy spectrum on large scales.

\subsubsection{Kinetic energy spectrum}
\label{subsection-kinetic-energy-spectrum}

The \OneD \KPZ equation for the profile ${h(t,x)}$ corresponds to the Burgers equation for ${\partial_x h(t,x)}$, which models a randomly stirred fluid. Consequently, in analogy with hydrodynamics, the kinetic energy density of the \OneD \KPZ dynamics can be defined as
\begin{align}
 E = \moy{\left(\nabla h\right)^2} = \int_{\omega,p} p^2 C(\omega,p) \, .
\end{align}
We introduce the kinetic energy spectrum as
\begin{align}
 \bar{R}(p) = \frac{p^{2}}{\pi} \int_0^\infty C(\omega,p) \text{d}\omega \, ,
 \label{eq:def_kinetic_spectrum}
\end{align}
such that $E = \int_{p} \bar{R}(p)$. It can be interpreted as the amount of kinetic energy contained in the Fourier \mbox{mode $p$}. $\bar{R}$ is also related to the derivative of the interface width (equal-time correlation function) as
\begin{align}
 \bar{R}(r) = \int_{{p}} \text{e}^{\I {p}{r}} \bar{R}(p) & = - \nabla^2 C(\tau=0,r) \nonumber \\
 & = \frac{1}{2} \nabla^2 {W}(\tau=0,r) \, .
 \label{eq:kinetic_to_roughness}
\end{align}

This function is precisely the quantity that has been studied analytically and numerically in \cite{agoritsas_2012_FHHtri-analytics,agoritsas_2012_FHHtri-numerics}. In these studies, the complete time-evolution of ${\bar{R}(t,r)}$ was investigated, starting from the `sharp-wedge' initial condition for the \OneD \KPZ equation. $\bar{R}(r)$ can be normalised as ${\bar{R}(r) \coloneqq [\widetilde{D}(\xi)/\xi] \, E_\xi(r/\xi)}$ with $\widetilde{D}(\xi)$ defined through $\int_r \bar{R}(r) = \widetilde{D}$. The normalisation and the shape of the correlator are characterised by ${\widetilde{D}(\xi)}$ and $E_\xi(x)$ respectively. Note that a similar normalisation yields in Fourier space ${\bar{R}(p) = \widetilde{D}(\xi) E_\xi(p \xi)}$ with ${E_\xi(q =0)=1}$. It was pointed out in \cite{agoritsas_2012_FHHtri-analytics,agoritsas_2012_FHHtri-numerics} that the function $E_\xi(x)$ is weakly dependent on $\xi$ as long as $\xi$ is not too large. More precisely, it was found that ${E_\xi = R_{\xi=1}}$ and ${\widetilde{D} = 1}$ in the limit ${\xi \rightarrow 0}$ and it was assumed that ${E_\xi \cong R_{\xi=1}}$ for small (but finite) values of $\xi$. In the opposite limit of very large $\xi$ the specific shape of ${E_{\xi}}$ remains an open issue. $\widetilde{D}(\xi\gg 1)$ as well as the crossover from $\xi\ll 1$ will be discussed in \sref{section-discussion-Dtilde}.

We have computed $\bar{R}$ for various values of $\xi$. The result of our calculation is shown in \fref{fig:rbar}. Since our solution of \Eq{eq:flow} for $f_k^X$ is numerical, it is restricted to a finite range of momenta and frequencies, in particular $\omega_1 \leq \omega \leq \Lambda^2$. For this reason, the frequency integral of \Eq{eq:def_kinetic_spectrum} must be split into three parts to be computed separately,
\begin{align}
  \bar{R}(p) = & \bar{R}_1(p) + \bar{R}_2(p) + \bar{R}_3(p) \nonumber \\
  = & \frac{p^2}{\pi}\Bigg[ \int_0^{\omega_1} + \int_{\omega_1}^{\Lambda^2}  + \int_{\Lambda^2}^\infty \Bigg] \, C(\omega,p)  \text{d}\omega \, .
  \label{eq:split_R}
\end{align}

First, the range $\omega_1 \leq \omega \leq \Lambda^2$ corresponds to the frequency range where numerical data for the dimensionful correlation function are available (see \aref{sec-appendix-numerics}), and the integral in $\bar{R}_2$ is computed numerically. Secondly, in the range $\omega \ge \Lambda^2$, the correlation function can be replaced by its bare form (obtained by inserting the initial conditions $f_\Lambda^X$ \Eq{eq:initial}, instead of $f_k^X$ in \Eq{eq:c_from_f}),
\begin{align}
 C(\omega,p) \cong \frac{2 R_\xi(p)}{\omega^2+p^4} \, ,
 \label{eq:bare_form}
\end{align}
since the high frequency sector is determined by the beginning of the \RG flow. The frequency integration in $\bar{R}_3$ is then performed analytically. Note that because of the exponentially decaying noise correlator, $\bar{R}_3$ is negligible compared to the other two parts of $\bar{R}$.

Third, in the range $\omega\leq \omega_1$, the fixed point is reached, and for $p\ll \Lambda$, the scaling form {of $C$ \Eq{eq:scaling_g}, can be inserted into the integral of ${\bar{R}_1}$,
\begin{align}
 \bar{R}_1(p) = \frac{D_\xi p^{-3/2}}{\pi \nu_\xi^2} \int_0^{\omega_1} \hat{F}\left(\frac{\omega}{\nu_\xi p^{3/2}}\right) \, \text{d}\omega \, .
\end{align}
The change of variables $\hat{x} = \omega/(\nu_\xi p^{3/2})$, and the exponent identity \Eq{eq:eta2chi} finally provides
\begin{align}
 \bar{R}_1(p)= \frac{D_\xi}{\pi \nu_\xi} \int_0^{\frac{\omega_1}{\nu_\xi p^{3/2}}} \hat{F}(\hat{x}) \, \text{d}\hat{x} \, .
 \label{eq:rbar1}
\end{align}
This expression is strictly valid for ${p\ll 1/\xi}$ (or ${p\ll\Lambda}$ when $\xi=0$). It can however be safely used for any $p$ because ${\bar{R}_1}$ is negligible when compared to ${\bar{R}_2}$ at larger values of $p$, see \fref{fig:rbar}.

As shown in the previous section, the integrand on the right-hand side of \Eq{eq:rbar1} is a universal quantity, which can be computed from the solution of the \RG fixed point alone. In particular when $p\rightarrow 0$ its integral becomes a universal constant. The pre-factor of \Eq{eq:rbar1} $D_\xi/(\pi\nu_\xi)$ is the limit ${D(k\rightarrow 0)/(\pi \nu(k\rightarrow 0))}$. The dependence of ${D(k)/(\pi\nu(k))}$ on the \RG scale $k$ drops out at small values of $k$ since the time-reversal symmetry is restored and yields $\eta_D = \eta_\nu$ \footnote{Note that setting ${f_k^\lambda=1}$ (see \sref{section-approximation}) is essential here. Without this approximation we only get $\eta_D \cong \eta_\nu$ and the limit ${D(k\rightarrow 0)/(\pi \nu(k\rightarrow 0))}$ is not finite.}. As manifest in \Eq{eq:expoflow}, its computation requires the entire solution of the \RG flow equations. This pre-factor is hence a non-universal quantity that depends on the value of $\xi$, as well as the specific form of $R_\xi$.

Two components of ${\bar{R}}$  are plotted in \fref{fig:rbar} for a representative value of $\xi = 1.6$. Note that ${\bar{R}_1}$ and ${\bar{R}_2}$ add up so that ${\bar{R}}$ is a constant at small $p$ (within a $0.16\%$ of relative accuracy). This is a consistency check of our calculation since the integrand on the {right-hand side} of \Eq{eq:rbar1} is computed once for all values of $\xi$ from the universal scaling form at $\xi=0$. We have checked that the crossover from ${\bar{R}_1}$ to ${\bar{R}_2}$ dominating ${\bar{R}}$ (at ${p\cong 2 \cdot 10^{-5}}$ on \fref{fig:rbar}) can be sent to arbitrarily small values of $p$ by taking $\omega_1$ small enough. At large $p$, the bare form \Eq{eq:bare_form} can be inserted in \Eq{eq:def_kinetic_spectrum} and provides $\bar{R}(p\gg 1/\xi) \cong R_\xi(p)$. This result is consistent with the prediction ${E_{\xi} \approx R_{\xi=1}}$ at small but finite $\xi$ discussed in \cite{agoritsas_2012_FHHtri-analytics,agoritsas_2012_FHHtri-numerics}, and suggests furthermore that in the stationary state ${\bar{R}(p)/\widetilde{D}(\xi)}$ differs from the microscopic ${R_{\xi}(p)}$ only for spatial modes ${p < 1/\xi}$.

\begin{figure}[tb]
\includegraphics[width=\columnwidth]{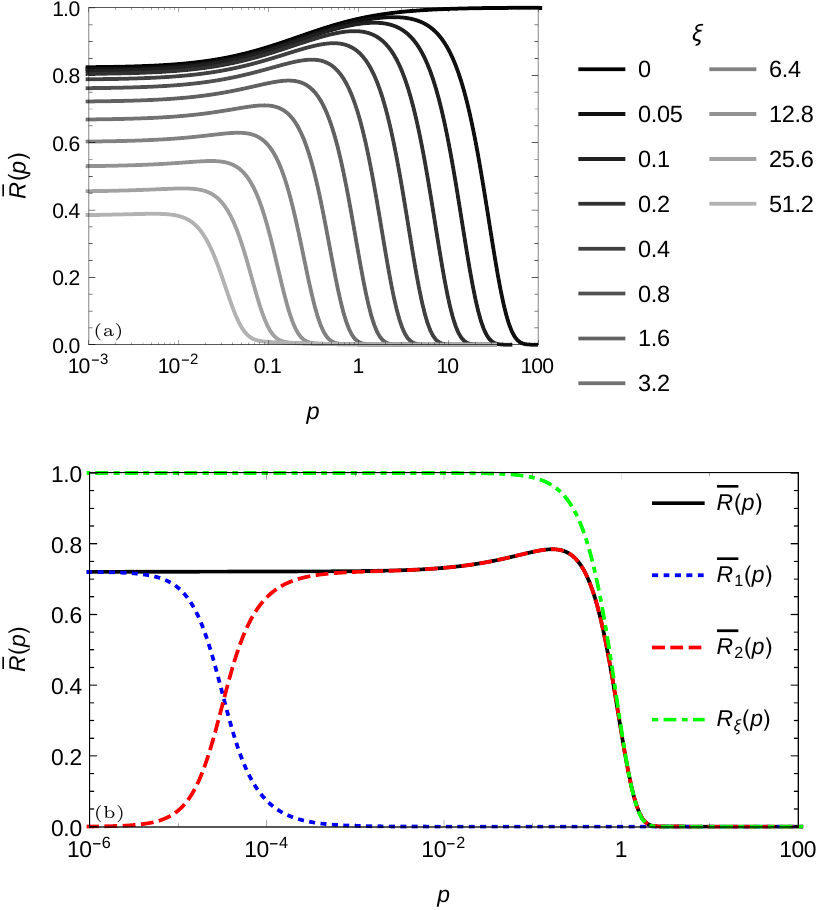}
\caption{(Colour online) Kinetic energy spectrum ${\bar{R}}$ - (a): ${\bar{R}}$ for different values of $\xi$. (b): ${\bar{R}}$ (solid, black), as well as its two dominating contributions [see \Eq{eq:split_R}], ${\bar{R}_1}$ (dotted, blue) and ${\bar{R}_2}$ (dashed, red) and the microscopic noise correlation function $R_\xi(p)$ (dash-dotted, green) for a representative value of $\xi=1.6$. The \UV physics is well captured by the bare noise correlator $R_\xi$ while the universal \KPZ physics (given by $\bar{R}_1$) emerges in the \IR domain.}
\label{fig:rbar}
\end{figure}

\subsubsection{Comparison between the \texorpdfstring{\FRG}{{NP-FRG}} predictions and previous numerical results}
\label{section-NPRG-comparison-with-FHHtrinum}

The \OneD \KPZ equation with correlated noise of range $\xi$ was studied in \cite{agoritsas_2010_PhysRevB_82_184207,agoritsas-2012-FHHpenta,agoritsas_2012_FHHtri-analytics,agoritsas_2012_FHHtri-numerics}. In particular direct numerical simulations were performed in \cite{agoritsas_2012_FHHtri-numerics}, where the Fourier transform of the kinetic energy spectrum, ${\bar{R}(r)}$ (denoted ${\bar{R}_{\text{sat}}(y)}$ in \cite{agoritsas_2012_FHHtri-numerics} and plotted in Fig.~6 therein) has been computed for different values of $\xi$. 

\begin{figure}[ht]
\includegraphics[width=\columnwidth]{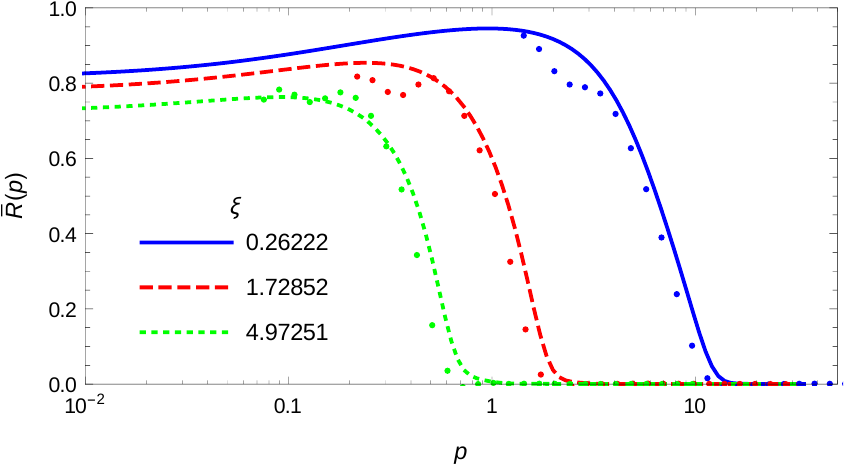}
\caption{(Colour online) Comparison of the \FRG calculation with direct numerical simulations - The kinetic energy spectrum ${\bar{R}}$ is computed for $\xi=0.2622$, $\xi_\tau=0.0579$ (blue, solid), $\xi = 1.7285$, $\xi_\tau = 1.3428$ (red, dashed) and $\xi=4.9725$, $\xi_\tau=7.8134$ (green, dotted). The lines show the \FRG result and the circles are the results of \cite{agoritsas_2012_FHHtri-numerics}.}
\label{fig:num_frg}
\end{figure}

The numerical simulations of \cite{agoritsas_2012_FHHtri-numerics} were achieved by sampling a noise with Gaussian statistics, computing the corresponding \KPZ time evolution and averaging at the end. A different system of units than ours was  used, see \aref{sec-appendix-scalings}. The correlation length was kept fixed and different values of the diffusion coefficient $\nu$ were considered. Moreover, a fixed correlation time $\xi_\tau'$, as well as a different form for ${R_{\xi}}$, were used for reasons of numerical stability  \cite{agoritsas_2012_FHHtri-numerics}.

The results of \cite{agoritsas_2012_FHHtri-numerics} can be easily converted to our units. The end result is that the variation of $\nu$ turns into a linked variation of ${\xi = \xi' D \lambda^2/\nu^3}$ and ${\xi_\tau = \xi_\tau' D^2 \lambda^4/\nu^5}$ (the parameters used in the simulation yield ${\xi_\tau \cong 0.539 \, \xi^{5/3}}$) and that the noise correlation function is well approximated by
\begin{align}
 R_{\xi,\xi_\tau}(\omega,p) = R'_\xi(p) R'_{\xi_\tau}(\omega) \, ,
 \label{eq:r_xi_xitau}
\end{align}
with the momentum and frequency correlator  being both given by
\begin{align}
 R'_l(y) = 9 \frac{\left[\text{sinc}\left(\frac{l y}{2}\right)\right]^8}{\left[2+\cos\left(l y \right) \right]^2} \, .
 \label{eq:r_xi_num}
\end{align}

We computed the \RG flows with these initial conditions for $f_\Lambda^D$ and different values of $\xi$ (and ${\xi_\tau \cong 0.539 \, \xi^{5/3}}$) and determined the corresponding kinetic energy spectra ${\bar R}$. \fref{fig:num_frg} shows a comparison of the \FRG results with the numerical results. Their quantitative agreement is very satisfactory. Note that the inclusion of a correlation time $\xi_\tau \neq 0$ explicitly breaks Galilean invariance \eq{eq:galilei} \cite{Medina1989a,katzav2004,Fedorenko2007,Strack2014a,Song2016b}. It is remarkable that this does not seem to affect the large-distance physics. Such a robustness of Galilean invariance was pointed out in \cite{Berera2007,Wio2010a,Wio2010b} (see \cite{Wio2016} for an overview). From an \RG point of view, this suggests that Galilean invariance is emergent like the time-reversal symmetry. We cannot confirm this statement within the \NLO approximation used in the present work since the breaking of Galilean symmetry by the initial condition $f_\Lambda^D = R_{\xi,\xi_\tau}$, generates violations of Ward identities for higher-order vertex functions, which are neglected at this order. In particular, the induced modification of the flow equation of $\hat g_k$ cannot be computed within \NLO approximation, such that  the  \RG flow of the theory is constrained to preserve the identity exponent $\chi+z=2$ at the \IR fixed point. However, the presence of temporal correlations do not affect the results found in \cite{agoritsas_2010_PhysRevB_82_184207,agoritsas-2012-FHHpenta,agoritsas_2012_FHHtri-analytics,agoritsas_2012_FHHtri-numerics} (where the full time evolution is considered) in any noticeable way when compared to results obtained in a Galilean invariant set-up. This strongly suggests that the \IR physics is indeed Galilean invariant.

\subsubsection{Crossover in the amplitude of the kinetic energy}
\label{section-discussion-Dtilde}

As already mentioned, the stationary kinetic energy spectrum tends to a constant as $p\rightarrow 0$,
\begin{align}
 \widetilde{D}(\xi) = \bar{R}(p=0) = \frac{D_{\xi}}{\pi \nu_{\xi}}  \int_0^\infty \hat{F}(\hat{x}) \text{d}\hat{x} \, . 
 \label{eq:dtilde_defint}
\end{align}
This constant can be related to the amplitude of the equal-time correlation function which, in the stationary state, takes the form
\begin{align}
 {W}({\tau=0},r) = {\widetilde{D}} \left|r\right|_\xi \, ,
 \label{eq:roughness_dtilde}
\end{align}
where $\left|\cdots\right|_\xi$ is a rounded (on scales given by $\xi$) absolute value \cite{agoritsas_2010_PhysRevB_82_184207,agoritsas-2012-FHHpenta,agoritsas_2012_FHHtri-analytics,agoritsas_2012_FHHtri-numerics}. This means that for $\xi=0$, one simply has $W(0,r) = \left|r\right|$. When $\xi$ is increased, the kink of the absolute value becomes smooth and the slope of $W(0,r)$ at large $r$ decreases. It is clear from \Eq{eq:roughness_dtilde} that ${\widetilde{D}}$ is observable on large scales. \Eq{eq:dtilde_defint} shows that it contains a $\xi$-dependent factor that can be extracted from our calculation. The result is shown in \fref{fig:dtilde}.

\begin{figure}[ht]
\includegraphics[width=\columnwidth]{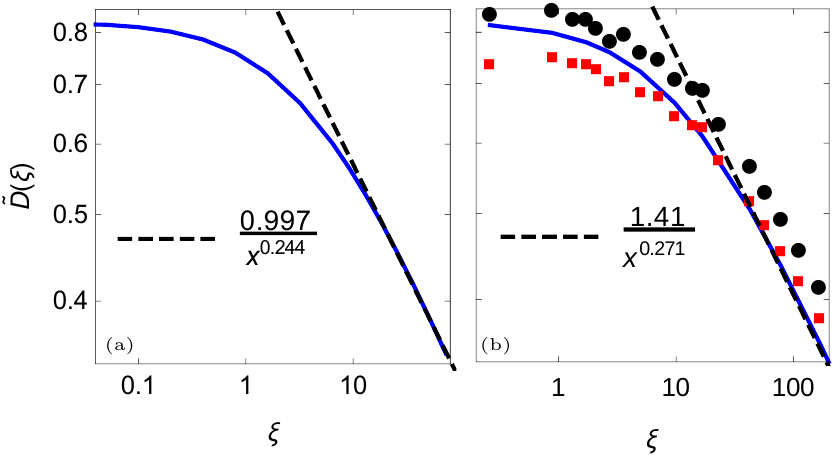}
\caption{(Colour online) Non-universal amplitude $\widetilde{D}$, in log-log scale, from \FRG and from numerical simulations - Plot (a) shows $\widetilde{D}$ computed with {\it spatial} noise correlations \Eq{eq:corr-xi}, with the \FRG and for different values of $\xi$ (blue, solid), as well as a power-law fit of the tail of the data (for $\xi\geq 18$) (black, dashed). Plot (b) shows $\widetilde{D}$ computed with {\it spatio-temporal} correlations \Eq{eq:r_xi_xitau}, with the \FRG for different values of $\xi$ and $\xi_\tau \cong 0.539 \, \xi^{5/3}$ (blue, solid) with its power-law fit (for $\xi \geq 43$) (black, dashed) as well as two different estimations of $\widetilde{D}$ from the numerical computation of \cite{agoritsas_2012_FHHtri-numerics} (black circles and red squares). The vertical axes of the two plots are scaled in the same way. The horizontal axis are not. The decay of $\widetilde{D}(\xi)$ with $\xi$ is a large-scale signature of the microscopic correlation length [see \Eq{eq:roughness_dtilde}]. It can be observed experimentally by varying the coupling $g=\lambda^2D/\nu^3$.}
\label{fig:dtilde}
\end{figure}

The behaviour of the amplitude ${\widetilde{D}}$ was previously addressed in \cite{agoritsas-2012-FHHpenta,agoritsas_2012_FHHtri-analytics,agoritsas_2012_FHHtri-numerics,phdthesis_Agoritsas2013}, fixing $\xi'$ and varying the diffusion coefficient $\nu$ within the original \KPZ equation \Eq{eq:kpz} (see \sref{section-NPRG-comparison-with-FHHtrinum}). A crossover was predicted analytically at $\xi \cong 1$, separating two limiting behaviours: at small values of the correlation length (${\xi\ll 1}$), a saturation of the amplitude to one ${\widetilde{D} \approx 1}$ \cite{Sasamoto2010a,Calabrese2011,Corwin2012a}, and in the opposite limit (${\xi \gg 1}$)  a decay as ${\widetilde{D} \sim \xi^{-1/3}}$. Note that the large-$\xi$ prediction relies on the existence of an optimal trajectory in the language of the directed polymer, as introduced in \cite{agoritsas_2012_FHHtri-analytics} and discussed more recently in \cite{agoritsas_lecomte_2016_scalings-GVM}. A key ingredient is to assume that the roughness of the polymer end-point free energy takes the form \Eq{eq:roughness_dtilde}. The recent Bethe ansatz analysis proposed in \cite{Dotsenko2016} also yields the same behaviour of $\widetilde{D}$ in the regime ${\xi\gg1}$, under the assumption of a $1$-step replica symmetry breaking in the replica description.

As for the behaviour of $\widetilde{D}$ at intermediate values of $\xi$, although no exact expression is available yet, two independent predictions have been obtained in the form of an implicit equation, ${\widetilde{D}^{\gamma} \propto (4/\xi)^{\gamma/3} (1 - \widetilde{D})}$, with ${\gamma \in \lbrace 3/2,6 \rbrace}$ and numerical pre-factors of order $10$ \cite{agoritsas_2012_FHHtri-analytics}, invoking in particular a \GVM computation with a full replica-symmetry-breaking \cite{agoritsas_2012_FHHtri-analytics,agoritsas_lecomte_2016_scalings-GVM}. These analytical predictions are qualitatively consistent with numerical measurements either on a directed polymer on a discrete lattice \cite{agoritsas-2012-FHHpenta} or in a direct numerical integration of the continuous \KPZ equation \cite{agoritsas_2012_FHHtri-numerics}. At last, we mention that ${\widetilde{D}}$ corresponds to the `fudging' parameter $f$ in \cite{agoritsas_2012_FHHtri-analytics,phdthesis_Agoritsas2013}, which coincides in fact with the full-replica-symmetry-breaking parameter in the \GVM computations presented in \cite{agoritsas_2010_PhysRevB_82_184207}.

We determined the non-universal amplitude ${\widetilde{D}}$ within the \FRG framework, both for a microscopic noise with spatial correlations \Eq{eq:corr-xi}, and different values of $\xi$, and for spatio-temporal correlations \Eq{eq:r_xi_xitau}, for different values of $\xi$ and correlation time set to ${\xi_\tau \cong 0.539 \, \xi^{5/3}}$. The two curves look very similar but differ quantitatively. By construction, the two saturation values at ${\xi=0}$ are the same. The results are displayed in \fref{fig:dtilde} alongside the data of Fig.~14 of \cite{agoritsas_2012_FHHtri-numerics} converted to our system of units. First, the crossover with $\xi$ and  the existence of two limiting regimes at large and small $\xi$ are recovered. We find qualitative agreement  with the corresponding analytical predictions: At $\xi=0$, the value is ${\widetilde{D}(0) \cong 0.82}$, and  at large ${\xi}$, the estimated power law  is ${\widetilde{D}(\xi \gg 1) \cong 1/\xi^{0.24}}$ for spatial correlations and ${\widetilde{D}(\xi \gg 1) \cong 1.4/\xi^{0.27}}$ for spatio-temporal correlations. Second, the \FRG results for spatio-temporal correlations are compared with the results from direct numerical simulations. The agreement is very precise for all values of $\xi$. Let us emphasise that this is a remarkable feature, since  $\widetilde{D}$ is a non-universal quantity, which hence depends on all the microscopic details. This shows that they are reliably captured by the \FRG flow.

The discrepancy between our value ${\widetilde{D}(0)\cong 0.82}$ and the analytical result ${\widetilde{D}(0) = 1}$ \cite{Sasamoto2010a,Calabrese2011,Corwin2012a} can possibly be attributed to the order of the approximation used (\NLO). The agreement would probably be improved at the next (\SO) order. See \sref{section-approximation} for a detailed discussion of the approximations involved here. Note that the method used to estimate $\widetilde{D}$ from the numerical data \cite{agoritsas_2012_FHHtri-numerics} is known to underestimate $\widetilde{D}$ and indeed ${\widetilde{D}(\xi=0)<1}$ in the data of \fref{fig:dtilde}. For the decay exponent, the observed difference could be a hint to the fact that for large $\xi$, one enters the regime of Burgers turbulence, which is dominated by shocks \cite{Bec2007a}. Hence, it is not clear whether the exponent values found ($-0.24$ and $-0.27$) are an artefact of the approximation scheme or not. The exploration of this regime within the \FRG formalism is beyond the scope of the present study.

\section{Connection with experiments}
\label{section-connection-exp}

The results presented here are experimentally accessible for systems that can be considered as continuous on scales smaller than $\xi'$,  that is $\xi' \gg a$ where $a$ is the microscopic lattice size. ${\widetilde{D}}$ provides an easily accessible observable because it can be measured on large scales [see \Eq{eq:roughness_dtilde}]. Although the detailed form of the noise may be hard to control experimentally, it is reasonable to assume (when $\xi' \gg a$) that there exists a small non-zero noise correlation length, $\xi'\neq 0$. In the original system of units,
\begin{align}
 \widetilde{D}'(\xi',D,\nu,\lambda) = \frac{D}{\nu} \widetilde{D}\left(\xi' \frac{\lambda^2 D}{\nu^3}\right) \, .
 \label{eq:dtilde_exp}
\end{align}
See \aref{sec-appendix-scalings} and \Eqs{eq:unit_conversion} for the details of the conversion. The behaviour ${\widetilde{D}}$ (\fref{fig:dtilde}) can be probed by varying $\nu$, $\lambda$ or $D$ instead of $\xi'$. Assuming that $\xi'$ is fixed, the control parameter becomes $g=\lambda^2 D/\nu^3$. If $g$ can be varied over one or two decades and if $\xi'$ is not too small, then a decrease of $\widetilde{D}'(\xi',D,\nu,\lambda)$ should be observable as $g$ is increased. Note that an extended discussion, on the different experiments in which these different predictions on $\bar{R}(p)$ and ${\widetilde{D}(\xi)}$ could be tested, is given in Sec.~VII. of \cite{agoritsas_2012_FHHtri-analytics}, using the units system recalled in \aref{sec-appendix-scalings}.

\section{Conclusion and perspectives}
\label{section-conclusion-perspectives}

We have used the \FRG to determine the full momentum dependence of the stationary two-point correlation function of the stochastic \KPZ equation with microscopic noise correlated at a finite spatial scale $\xi$. We have resolved the non-universal features at scales smaller than $\xi$ as well as the universal scaling regime on large scales. We have shown (within our approximation scheme) that the universal physics is governed by the presence of a fully attractive \RG fixed point and does not depend on the microscopic noise correlation length, \fref{fig:rg_flow}. This implies that the time-reversal symmetry \eq{eq:time_reversal} that is broken at the microscopic level when ${\xi > 0}$ is emergent on large scales.

We computed the kinetic energy spectrum of the stationary \KPZ dynamics, which is a single-time observable that is related to the interface width through two space derivatives and a Fourier transform. Both the $\xi$-dependent \UV physics and the universal \IR physics are visible in \fref{fig:rbar}. Our results extend previous numerical simulations \cite{agoritsas_2012_FHHtri-numerics} to values of momenta that were not accessible before, \fref{fig:num_frg}. Finally we provide an experimentally accessible observable ${\widetilde{D}}$ and compute its dependence on $\xi$, \fref{fig:dtilde}. Our results are in good agreement with the numerical results of \cite{agoritsas_2012_FHHtri-numerics} as well as with other analytical results \cite{agoritsas_2010_PhysRevB_82_184207,agoritsas-2012-FHHpenta,agoritsas_2012_FHHtri-analytics,agoritsas_2012_FHHtri-numerics,Dotsenko2016} (although qualitatively). Calculation at the next-order approximation would certainly  improve the results. In particular they would help to settle the discrepancy in the obtained value of the decay exponent of $\widetilde{D}(\xi\gg 1)$ with respect to the result of \cite{agoritsas_2010_PhysRevB_82_184207,agoritsas-2012-FHHpenta,agoritsas_2012_FHHtri-analytics,agoritsas_2012_FHHtri-numerics}. In this respect, an experimental (or alternative) determination would be desirable.

An interesting direction would be to investigate $3$-point correlation functions and extend this study to larger values of $\xi$, where the regime of \OneD Burgers turbulence, with an energy cascade developing, could be investigated. This is left for future work.

\begin{acknowledgments}
S.~M. and E.~A. acknowledge financial support from the Swiss National Science Foundation. E.~A. acknowledges additional financial support from ERC grant ADG20110209. V.~L.~acknowledges support from the ERC Starting Grant 680275 MALIG and the ANR-15-CE40-0020-03 Grant LSD. The authors also thank  Nicol\'{a}s Wschebor, Natalia Matveeva and Eiji Kawasaki for useful discussions.
\end{acknowledgments}

\appendix

\section{Different units}
\label{sec-appendix-scalings}

In this appendix we detail the change of units relating \Eqs{eq:kpz} and \eq{eq:kpz_du}, and the units used in the numerical simulations \cite{agoritsas_2012_FHHtri-numerics}. \Eq{eq:kpz} is defined in terms of dimensionful parameters, $\lambda$, $\nu$, $D$ and $\xi'$. Space, time and fields can be rescaled. Since the dimensions of $\eta'$ and $h'$ are related, three parameters can be set to one by an appropriate choice of units. We choose to keep $\xi$ as the unique free parameter. The following rescaling,
\begin{align}
 & t' = \frac{1}{\nu} \left( \frac{\nu^3}{D \lambda^2}\right)^{2} \, t \, , && x' = \frac{\nu^3}{D \lambda^2} \, x \, , && \nonumber \\
 & h' = \frac{\nu}{\lambda} \, h \, , && \eta' = \frac{\nu^2}{\lambda} \left( \frac{D \lambda^2}{\nu^3}\right)^{2} \, \eta \, ,
 \label{eq:unit_conversion}
\end{align}
converts the dimensionful quantities of \Eq{eq:kpz} (noted here with a $'$) to the dimensionless ones of \Eq{eq:kpz_du}. The remaining parameter is $\xi = \xi' \lambda^2 D/\nu^3$. In particular, $\widetilde{D}'$ takes the form given by \Eq{eq:dtilde_exp} and the dimensionful kinetic energy spectrum becomes
\begin{align}
 \bar{R}'(p') = \frac{D}{\nu} \bar{R}\left(\frac{\nu^3}{D \lambda^2}p'\right) \, .
 \label{eq:dim_R}
\end{align}

In the numerical simulations  \cite{agoritsas_2012_FHHtri-numerics}, the \KPZ equation \Eq{eq:kpz} is written in terms of the parameters $T$, $c$ and $\Delta$ with the correspondence
\begin{align}
 \nu = \frac{T}{2c} \, ,&& \lambda = -\frac{1}{c} \, , && D = \frac{\Delta}{2} \, .
 \label{eq:polymer_to_kpz}
\end{align}
These parameters are inherited from the exact mapping between a thermally equilibrated \OneD elastic interface in a short-range correlated disorder and a directed polymer growing in a two-dimensional disordered energy landscape \cite{huse_henley_fisher_1985_PhysRevLett55_2924,HalpinHealy1995}. From there, the \KPZ equation is recovered by noting that the polymer-endpoint free energy evolves (with the polymer length) according to the \KPZ equation with `sharp-wedge' initial conditions and the \KPZ parameters ${\lbrace \nu, \lambda,D,\xi' \rbrace}$ obtained from  \Eqs{eq:polymer_to_kpz} \cite{huse_henley_fisher_1985_PhysRevLett55_2924,bouchaud_mezard_parisi_1995_PhysRevE52_3656}. In the language of the elastic interface, $c$ is the elastic constant, $T$ is the temperature, and $\Delta$ is the amplitude of the microscopic disorder two-point correlator. In addition, $\xi'$ is defined as the disorder correlation length, but it can alternatively correspond to the typical thickness of the interface \cite{agoritsas_2012_ECRYS2011}. Note that in these units, the two opposite limits of ${\xi \to 0}$ and ${\xi \to \infty}$ translate respectively into the limits of `high temperature' (${T \gg T_c}$) and `low temperature' (${T \ll T_c}$), with a characteristic crossover temperature ${T_c(\xi')=(\xi' c D)^{1/3}}$.

\section{Principle of Minimum Sensitivity}
\label{sec-pms}

The cut-off matrix \eq{eq:cut_off_matrix} contains an arbitrary parameter $\alpha$. In principle, the end result of a calculation should not depend on $\alpha$. However, any approximation introduces a spurious dependence on the cut-off matrix. An optimal value for $\alpha$ can be determined according to the principle of minimum sensitivity \cite{Stevenson1981,Canet:2002gs}, which leads to extremising the quantity computed with respect to $\alpha$.

The present calculation turns out to be relatively insensitive to $\alpha$. We hence determined its optimal value  using a single observable, $\widetilde{D}(0)$, defined in \Eq{eq:dtilde_defint}, and we used the same optimal value for all $\xi$. This procedure is illustrated in \fref{fig:pms}, which shows the variation of $\widetilde{D}(0)$ with respect to $\alpha$. A third-order polynomial fit  yields $\alpha_{\text{opt}} \cong 38.82$.

\begin{figure}[ht]
\centering
\includegraphics[width=\columnwidth]{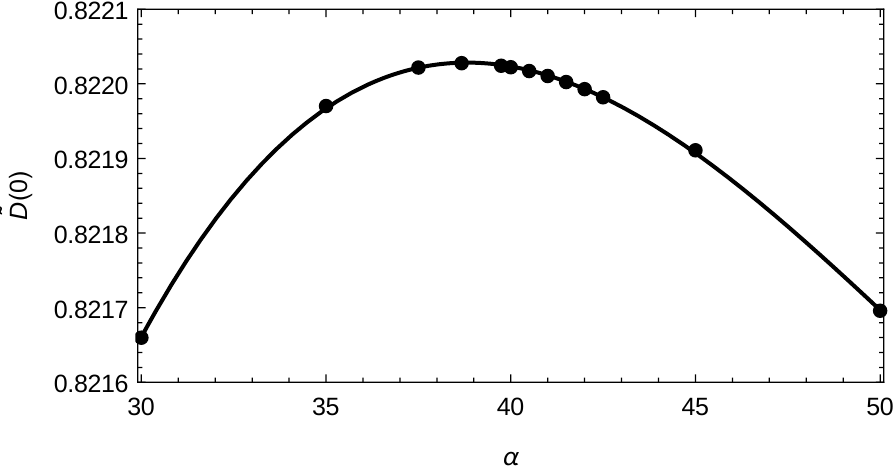}
\caption{Principle of minimum sensitivity - The circles represent the \FRG results for $\widetilde{D}(0)$ as a function of $\alpha$. The line is a third order polynomial fit.}
\label{fig:pms}
\end{figure}

Note that the variations of $\widetilde{D}(0)$ around its optimal value are small. The main source of error comes from the order of truncation (\NLO). The latter can be assessed by comparing the \FRG result $\widetilde{D}(0) \simeq 0.82$ for the \KPZ equation with uncorrelated noise ($\xi=0$), to the exact result $\widetilde{D}(0) = 1$, which hints at roughly a $20\%$ error.

\section{Numerical solution of the \texorpdfstring{\RG}{RG} flow equations}
\label{sec-appendix-numerics}

In this appendix, we give some details on the numerical solution of \Eqs{eq:flow} and \eq{eq:dimensionfull_flow}. Dimensionless quantities such as $\hat{f}_k^X$ and $\eta_X(k)$ are directly obtained from the numerical solution of \Eqs{eq:flow}. Dimensionful observables such as $C$ are then extracted from this solution. This procedure was introduced in \cite{Benitez2009,Benitez2012}, and is here generalised to frequency-dependent quantities.

\subsection{Solution of the dimensionless flow equations}

We follow the numerical scheme  used in \cite{kloss_canet_wschebor_2012_PhysRevE86_051124}. A large value of $\Lambda=200$ is chosen where the initial conditions \Eqs{eq:initial_rescaled} are set for each value of $\xi$. The dimensionless frequency and momentum are discretised into regular grids, $\hat{p}_i = \hat{p}_1 + \Delta \hat{p} \, (i-1)$ (with $i=1,N_p$) $\hat{\omega}_j = \hat{\omega}_1 + \Delta \hat{\omega} \, (j-1)$ (with $j=1,N_\omega$) and the functions $\hat{f}_k^X(\hat{\omega},\hat{p})$ are represented as $N_p\times N_\omega$ matrices (with only positive momenta and frequencies  since the solution of \Eq{eq:flow} only depends on the absolute values $\left|\hat{p}\right|$, $\left|\hat{\omega}\right|$). The values of $N_p$, $N_\omega$, $\Delta\hat{p}$, $\Delta\hat{\omega}$ as well as all the other numerical parameters that we use are specified in \tref{tb:num_parameters}.

A third order polynomial spline is used to compute $\hat{f}_k^X$ for momenta and frequencies that are not in the tabulated set of values. In particular, this spline is used to compute the derivatives in the linear part of the right-hand side of \Eq{eq:flow}. For ${\hat{p} \ge \hat{p}_{N_p}}$, the functions are approximated by power laws, ${\hat{f}_k^X(\hat{\omega},\hat{p}) = \hat{f}_k^X(\hat{\omega},\hat{p}_{N_p}) (\hat{p}/\hat{p}_{N_p})^\eta}$ with
\begin{align}
\eta = \frac{\ln[\hat{f}_k^X(\hat{\omega},\hat{p}_{N_p})]-\ln[\hat{f}_k^X(\hat{\omega},\hat{p}_{N_p}-0.2)]}{\ln(\hat{p}_{N_p})-\ln(\hat{p}_{N_p}-0.2)} \, , 
\end{align}
where $\hat{f}_k^X(\hat{\omega},\hat{p}_{N_p}-0.2)$ is computed with the spline.

To compute the non-linear integrals $\hat{I}^X$ on the right-hand side of \Eq{eq:flow}, the \NLO approximation for the frequency dependence is used [the replacement ${\hat{f}_k^X(\hat{\Omega},\hat{Q}) \rightarrow \hat{f}_k^X(0,\hat{Q})}$ for all configurations $\hat \Omega$ and $\hat Q$ inside the integrands $\hat{J}^X$, which are defined in \Eq{eq:dimensionfull_flow}]. This is exploited to perform the integration over the internal frequency analytically, see \cite{kloss_canet_wschebor_2012_PhysRevE86_051124} for detailed expressions. The integrals over the internal momentum $\hat{q}$ are computed with a Gauss-Legendre quadrature. Because of the insertion of the cut-off matrix and its scale-derivative [see \Eq{eq:wetterich}] the remaining integrand ${\int_{\hat{f}} \hat{J}^X(\hat{\omega},\hat{p},\hat{f},\hat{q})}$ is a smooth function of $\hat{p}$, $\hat{\omega}$ and $\hat{q}$. Moreover, the insertion of $k\partial_k \mathcal{R}_k$ imposes that  this integrand is exponentially suppressed for $\hat{q}$ and $\left|\hat{p}\pm\hat{q}\right| \gg 1$. Consequently, a coarser grid for $\hat{q}$ and a smaller range of internal momenta $\hat{q}<q_{\text{max}}$ can be used to compute the integral on $\hat q$ without loss of precision.

The lowering of the \RG scale is performed with an explicit Euler time stepping in the \RG time,
\begin{align}
 \hat{f}_{k(1-ds)}^X = \hat{f}_{k}^X - ds \, k\partial_k\hat{f}_k^X \, ,
\label{eq:split}
\end{align}
where $ds$ is the step size. This procedure is iterated until the cut-off scale is much smaller than all the dimensionful momenta that are considered, $s\rightarrow s_{\text{min}}$.

Finally, an additional procedure  is implemented to correctly resolve the momentum dependence of $\hat{f}_k^D$ for large $\xi$ at the beginning of the flow. Indeed, since $\Lambda$ is taken to be $200$, $\hat{\xi} = \Lambda \xi$ can be large even when $\xi$ is not. Then $\hat{f}_\Lambda^D(\hat{\omega},\hat{p}) = R_{\hat{\xi}}(\hat{p})$ decays exponentially at a scale given by $\hat{\xi}$. If treated directly, this would impose the choice of $\Delta \hat{p}$  very small and $N_p$ very large. 

On the other hand, at the beginning of the flow, when $k \cong \Lambda$, $\hat{g}_k = g/k$ is very small and the \RG flow equations are almost linear. Hence, lowering the \RG scale amounts to a rescaling. Thus, at the beginning of the flow, the renormalized forcing correlator is separated into two parts,
\begin{align}
\hat{f}_k^D(\hat{\omega},\hat{p}) = R_{\hat{\xi} \, k/\Lambda}(\hat{p})/D(k) + \delta \hat{f}_k^D(\hat{\omega},\hat{p}) \, ,
\end{align}
where only the first term has sharp momentum variations at the beginning of the flow. Its \RG flow  can be determined analytically, knowing $\eta_{D}(k)$. For the second term, the \RG flow starts with $\delta \hat{f}_k^D=0$. As long as $s = \ln(k/\Lambda)$ is close to zero, $\hat{f}_k^D$ is only weakly renormalized and $\delta \hat{f}_k^D$ remains small and smooth, and can be treated numerically. Its flow obeys an equation  obtained from \Eq{eq:flow} by substituting $\hat{f}_k^D \rightarrow \delta \hat{f}_k^D$ on the left-hand side and $\hat{f}_k^D(\hat{\omega},\hat{p}) \rightarrow R_{\hat{\xi}\, k/\Lambda}(\hat{p})/D(k) + \delta \hat{f}_k^D(\hat{\omega},\hat{p})$ inside $\hat{I}^X$. When a spline or interpolation is necessary in $\hat{I}^X$, this is only applied to $\delta \hat{f}_k^D$, and the analytical form of $R_{\hat{\xi}\,k/\Lambda}(\hat{p})/D(k)$ is used. Then the rest of the momentum integration is performed as outlined above and with the same momentum grid.

As the effective correlation length $\hat{\xi} k/\Lambda$, decreases with the \RG scale, $\hat{f}_k^D$ becomes smooth. We use the criterion that when $\hat{\xi} k/\Lambda \, \hat{p}_5 = k \xi \hat{p}_5 < 1$, $\Delta \hat{p}$ is small enough for a straightforward numerical solution, the splitting \Eq{eq:split} ceases and the flow of the whole function $\hat{f}_k^D$ is computed at once. This procedure is used in the same way in the frequency variables when $\xi_\tau$ is non zero.

\subsection{Solution of the dimensionful flow equations}
\label{sec-appendix-reconstr-dimensions-after-RG}

The dimensionful correlation function $C$ can be computed for arbitrary  momentum and frequency with a great accuracy and low computational cost by suitably using the solution of \Eqs{eq:flow}, instead of directly solving \Eqs{eq:dimensionfull_flow}, following \cite{Benitez2009,Benitez2012}. The idea is to compute the flow in two parts. For each dimensionful momentum $p$ and frequency $\omega$, the beginning of the flow from $k=\Lambda$ to $k=k_s(\omega,p)$ [with $k_s(\omega,p)$ specified below in \Eqs{eq:ks1} and \eq{eq:ks2}] is computed in the dimensionless representation with the procedure described previously, yielding $\hat f^X_{k_s}$. The dimensionful solution (at $k=k_s$) is then obtained through \Eqs {eq:rescalings},
\begin{align}
 f_{k_s}^X(\omega,p)  = X_{k_s} \hat{f}_{k_s}^X\left(\frac{\omega}{k_s^2 \nu_{k_s}},\frac{p}{k_s}\right) \, .
 \label{eq:frozen_flow}
\end{align}
The end of the flow from $k_s(\omega,p)$ to ${k = \Lambda \, \text{e}^{s_\text{min}}}$ is computed on a secondary dimensionful grid, yielding $f^X_{k\rightarrow 0}$. 
   
Here, we work with a logarithmic grid for the dimensionful momenta and frequencies: $p_i = \Lambda \text{e}^{dp(i-M_p)}$, with $i=1,M_p$ and $\omega_j = \Lambda^2 \text{e}^{d\omega(j-M_\omega)}$, with  $j=1,M_\omega$. The scale $k_s(\omega_j,p_i)$ is defined for each $(\omega_j,p_i)$ such that the corresponding dimensionless momentum $p_i/k_s$ or frequency $\omega_j/(k_s\nu_{k_s}^2)$ attains a large given value $\hat{p}_{n_p}$ or $\hat{\omega}_{n_\omega}$. The indices $n_p$ and $n_\omega$ are hence chosen close to the boundary of the grid (a few points before not to be affected by effects from the finiteness of the grid). More precisely, $k_s(\omega_j,p_i)$ is defined as
\begin{align}
  k_s(\omega_j,p_i) = \text{max}(k_1,k_2) \, ,
  \label{eq:ks1}
\end{align}
with
\begin{align}
   p_i = k_1 \hat{p}_{n_p} \, , && \omega_j = k_2^2 \nu_{k_2} \hat{\omega}_{n_\omega} \, .
   \label{eq:ks2}
\end{align}
Hence, for the second part of the flow on the dimensionful grid $k<k_s$, the condition $p_i\gg k$ is always satisfied. This choice is tailored to ensure that the flow on the dimensionful grid can be approximated in a simple and accurate way, such that all dimensionful momenta  values are decoupled. Indeed,  the presence of the derivative of the regulator $\partial_k R_k$, which is peaked at values $q\simeq k$, in the non-linear integrals on the right-hand side of \Eqs{eq:dimensionfull_flow} effectively cuts off  the internal momentum $q$ to values $q\lesssim k$. This yields that any $p_i$ computed on the dimensionful grid (satisfying $p_i\gg k$), also verifies $p_i\gg q$, and the integrand  $J_k^X(\omega_j,p_i,f,q)$ in \Eq{eq:dimensionfull_flow} can be Taylor expanded to leading order in powers of $q/p_i$. Hence, within this approximation, the flow equations become local in momentum space, \ie the integral takes the form $I^X_k(\omega_j,p_i) = L_{k1}(p_i)\int_q L_{k2}(\omega_j,q)$, with $L_{k1}(p)$ and $L_{k2}(\omega_j,q)$ two functions that can be extracted from the Taylor expansion of $J_k^X(\omega_j,p_i,f,q)$. The integral over the internal momentum  $q$ can be computed once (at every time step) for all values of $p_i$ using the dimensionless grid, such that the flow equations for the different momenta $p_i$ on the dimensionful grid are no longer coupled  together.

Let us unfold the practical sequence  to solve the \RG equations: Two independent and constant grids are defined, a dimensionful $(\omega \times p)$ grid and a dimensionless $(\hat \omega \times \hat p)$ one, on which the \RG flow is calculated [with \Eq{eq:flow}]. To describe the interplay between the two grids, we refer in the following to the (dimensionless) grid that is obtained by rescaling the dimensionful grid-points according to the rescaling procedure given in \Eq{eq:rescalings} as `the rescaled dimensionful grid'. As the grid-point values of the dimensionful grid stay constant, the values of the corresponding rescaled dimensionful grid grow according to Eq.\ (\ref{eq:rescalings}) when the scale $k$ decreases during the \RG flow. Moreover, we choose the value of $\Lambda$ such that
\begin{align}
  p_{M_p} < \Lambda \hat{p}_{n_p} \, , && \omega_{M_\omega} < \Lambda^2 \hat{\omega}_{n_\omega} \, ,
\label{eq:big_Lambda}
\end{align}
which ensures that, at the beginning of the flow, all tabulated rescaled dimensionful momenta $p_i/k$ and frequencies $\omega_j/(\nu(k) k^2)$ lie inside the dimensionless grid. This allows to solve the dimensionful flow equation \Eq{eq:dimensionfull_flow} at the beginning of the flow on this grid, by identifying the rescaled dimensionful grids points with the dimensionless ones.

As $k$ decreases, the grid-points of the rescaled dimensionful grid grow until they run out-of-range of the dimensionless grid one by one consecutively (in both, frequency and momentum direction). For convenience, we set ${n_p = N_p}$ and ${n_\omega = N_\omega}$ in the following discussion. The generalisation is evident. When a rescaled dimensionful momentum or frequency variable of a given grid-point hits the edge of the dimensionless grid, at the corresponding scale $k_s$, the values of the dimensionful flow functions $f_{k_s}^X(\omega,p)$ on that grid-point of the dimensionful grid are deduced according to \Eq{eq:frozen_flow}. Once outside the range of the dimensionless grid, the evolution of $f_{k}^X(\omega,p)$ (with $\omega$ and $p$ on the dimensionful grid) is much slower and allows to compute the flow in the dimensionful representation from $J_k^X(\omega_j,p,f,q)$, which is Taylor expanded to leading order in $q/p$, and all momenta decouple.

Finally, as the \RG time $s$ decreases in discrete steps, the time $s$ when a rescaled dimensionful grid-point hits the border of the dimensionless grid falls in general in between two discrete time steps $ds$. A linear interpolation is used to evaluate $\hat{f}_{k_s}^X$ and $\eta_X(k_s)$ in between the two surrounding time steps.

All the points of the dimensionful grid sequentially attain the edges of the dimensionless grid and all the rows and columns of the dimensionful grid for $f_{k}^X(\omega_j,p_i)$ progressively start running. In fact, the dimensionful flow effectively stops very rapidly because of the decoupling property outlined in \sref{subsection-two-point-correlator}, and even the zeroth order of the Taylor expansion in $q/p$ can be used.
 
Note also that the flow of $f_{k}^X(\omega,p)$ if either $\omega$ or $p$ is exactly zero cannot be computed on the dimensionful grid. However, the first values $p_1$ and $\omega_1$ can be rendered arbitrarily small by lowering further the \RG scale (carrying the numerical integration longer). On the other hand, the zero momentum or frequency properties are directly extracted from the dimensionless grid.
\begin{table}[ht]
\begin{tabular}{l l p{0.05\textwidth} l l}
\hline
Parameter & Value & & Parameter & Value \\
\toprule
 $\Delta\hat{p}$ & $0.128$ & & $\Delta\hat{\omega}$ & $0.25$ \\
 %\hline
 $N_p$ & $161$ & & $N_\omega$ & $121$ \\
 %\hline
 $\hat{p}_0$ & $10^{-10}$ & & $\hat{\omega}_0$ & $10^{-10}$ \\
% \hline
 $N_g$ & $30$ & & $q_{\text{max}}$ & $4$ \\
% \hline
 $ds$ & $0.00002$ & & $s_{\text{min}}$ & $-24.6$ \\
 %\hline
 $M_p$ & $300$ & & $M_\omega$ & $300$ \\
 %\hline
 $dp$ & $0.072$ & &  $d\omega$ & $0.090$ \\
 $n_p$ & $145$ & &  $n_\omega$ & $105$ \\
 \hline
 \end{tabular}
 \caption{Parameters for the numerical solutions of \Eqs{eq:flow} and \eq{eq:dimensionfull_flow}}
 \label{tb:num_parameters}
\end{table}

\section{Extraction of the scaling function and non-universal amplitudes}
\label{sec-appendix-extracting}

According to \sref{subsection-two-point-correlator}, the scaling functions associated with the two-point correlation function are the same (for all values of $\xi$) up to normalisations. To test this in the numerical solution, we first extracted the scaling function $\hat F$ at $\xi=0$ from the combination of \Eqs{eq:fixed_point_scaling} and \eq{eq:Gxi}. As shown in \cite{canet_2011_PhysRevE84_061128}, this function can be accurately represented by a family of fitting functions that are power laws of rational fractions
\begin{align}
 \hat{F}(\hat{x}) = \left[ \frac{\sum\limits_{i=1}^{n-1} A_{2i} \, \hat{x}^{2(i-1)}+ A_{2n-1} T^{6/7} \, \hat{x}^{2(n-1)}}{1+\sum\limits_{i=1}^n A_{2i-1} \, \hat{x}^{2i}} \right]^{7/6} \, .
 \label{eq:fit_formula}
\end{align}
$T$ and $A_i$ (with ${i=1,2n-1}$) are positive fitting parameters. Here, we use ${n=4}$ and denote by ${\hat F_{\text{fit}}(\hat{x})}$ the corresponding fitting function (extracted at ${\xi=0}$). We then numerically compute for each value of $\xi$ the scaling function $F_\xi^K$ obtained from the truncated two-point correlation function where momenta ${p>K}$ and frequencies ${\omega > K^{3/2}}$ are removed [see \Eq{eq:non_universal_scaling}]. We determine the non-universal normalisations ${\alpha_\xi^K}$ and ${\beta_\xi^K}$ defined by 
\begin{align}
F_{\xi}^K(x) = \alpha_\xi^K \, \hat{F}_{\text{fit}}(\beta_\xi^K x) \, ,
\label{eq:same_scaling}
\end{align}
as a function of $K$, using $\alpha_\xi^K$ and $\beta_\xi^K$ as fit parameters and keeping $\hat{F}_{\text{fit}}$ fixed.
   
We checked that when $K$ is small enough, $\hat F_{\text{fit}}$ indeed represents an appropriate scaling function for $F_\xi^K$ (extracted at $\xi\neq 0$). The values of the obtained fitting parameters  $\alpha_\xi^K$ and  $\beta_\xi^K$ are displayed in \fref{fig:same_scaling} as a function of $K$ for different values of $\xi$. Below a certain threshold of  $K$, these parameters become constants (independent of $K$) and the error of the fit is very small. It is clear from \fref{fig:same_scaling} that scale invariance emerges on large scales. Conversely, when $K$ is too large, there is no scale invariance and the fitting procedure fails.

\begin{figure}[t]
\begin{center}
 \includegraphics[width=0.99\columnwidth]{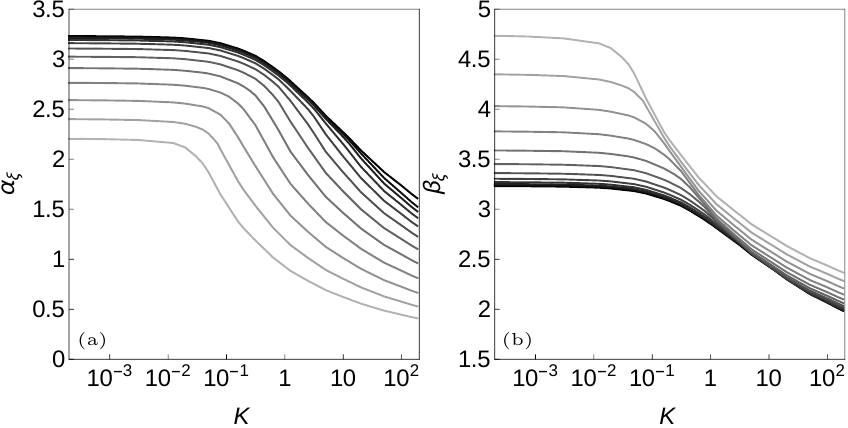}
\end{center}
\caption{Fitting parameters as functions of $K$: (a) $\alpha_\xi^K$ and (b) $\beta_\xi^K$ - The different levels of grey represent different values of $\xi$, with the same legend as in \fref{fig:rbar}. If enough of the small-scale physics $K\lessapprox \text{min}(1/\xi,\Lambda)$ is excluded, \Eq{eq:same_scaling} provides an adequate fitting form for all the scaling functions.}
\label{fig:same_scaling}
\end{figure}
\FloatBarrier

\end{document}